\documentclass[onecolumn,prl,aps,superscriptaddress,11pt]{revtex4-1}
\usepackage[T1]{fontenc}
\usepackage[latin9]{inputenc}

\usepackage{amsmath}
\usepackage{amssymb}
\usepackage{bbm}
\usepackage{braket}

\usepackage{graphicx}

\usepackage[colorlinks=true]{hyperref}  
\hypersetup{
    bookmarks=true,         
    unicode=false,          
    pdftoolbar=true,        
    pdfmenubar=true,        
    pdffitwindow=false,     
    pdfstartview={FitH},    
    pdftitle={Identifying topological order through unsupervised machine learning},    
    pdfauthor={Rodriguez-Nieva and Scheurer},     
    pdfsubject={},   
    pdfcreator={},   
    pdfproducer={}, 
    pdfkeywords={} {} {}, 
    pdfnewwindow=true,      
    colorlinks=true,       
    linkcolor=blue, 
    citecolor=blue,        
    filecolor=magenta,      
    urlcolor=blue,           
	breaklinks=true
}

\newcommand{\equref}[1]{Eq.~(\ref{#1})}

\newcommand{\figref}[1]{Fig.~\ref{#1}}

\newcommand{\refcite}[1]{Ref.~\onlinecite{#1}}

\newcommand{\diff}{\mathrm{d}}

\newcommand{\sect}[1]{\vspace{1em}{\bf #1}.~}

\newcommand{\be}{\begin{equation}}
\newcommand{\ee}{\end{equation}}
\newcommand{\bea}{\begin{eqnarray}}
\newcommand{\eea}{\end{eqnarray}}

\newcommand{\sign}{\,\text{sign}}
\renewcommand{\approx}{\simeq}

\renewcommand{\vec}[1]{\boldsymbol{#1}}

\renewcommand{\thefigure}{{\bf \arabic{figure}}}

\newcommand{\SuplDisc}[1]{{Supplementary Information}}
\newcommand{\SD}{Supplementary Discussion }

\begin{document}
\title{Identifying topological order through unsupervised machine learning}
\author{Joaquin F. Rodriguez-Nieva}
\email{jrodrigueznieva@fas.harvard.edu}
\affiliation{Department of Physics, Harvard University, Cambridge MA 02138, USA}

\author{Mathias S. Scheurer}
\email{mscheurer@g.harvard.edu}
\affiliation{Department of Physics, Harvard University, Cambridge MA 02138, USA}

\date{\today}

\begin{abstract}

The Landau description of phase transitions relies on the identification of a local order parameter that indicates the onset of a symmetry-breaking phase. In contrast, topological phase transitions evade this paradigm and, as a result, are harder to identify. Recently, machine learning techniques have been shown to be capable of characterizing topological order in the presence of human supervision. Here, we propose an unsupervised approach based on diffusion maps that learns topological phase transitions from raw data without the need of manual feature engineering. Using bare spin configurations as input, the approach is shown to be capable of classifying samples of the two-dimensional XY model by winding number and capture the Berezinskii-Kosterlitz-Thouless transition. We also demonstrate the success of the approach on the Ising gauge theory, another paradigmatic model with topological order. In addition, a connection between the output of diffusion maps and the eigenstates of a quantum-well Hamiltonian is derived. Topological classification via diffusion maps can therefore enable fully unsupervised studies of exotic phases of matter.

\end{abstract}





\maketitle

Machine learning (ML) techniques \cite{NielsensBook,GoodfellowBook,mehta2018Review} are exquisitely tailored to uncover structure in complex data and to efficiently describe it with a minimal number of parameters. Recent advances in this field, demonstrated in areas as diverse as computer vision and natural language processing, have motivated the application of ML to various problems in condensed matter physics, with the hope of mutual benefits. Of the many conceptually and practically valuable applications \cite{ConnectionToRG,Carleo602,Sarma,FuSelfLearning,QuantumStateTomography,YZYouGeometry,carleo2018constructing,InteractionDistance}, using ML to classify phases of matter and identify phase transitions has taken center stage \cite{RogerNature,MelkoFermions,PhysRevLett.118.216401,QuantumLoopTopography2,Confusion,Rand2DSystem,RandomSystems,yoshiokaDisorder,S1S1MappingSupervised,RealSpaceTopInvs,AdversarialNNs,iakovlev2018supervised,vargas2018extrapolating,RogerXY,wang2018machine,HuUnsupervised,Wetzel,WangUnsupervised,cristoforettiUnsupervised,TrebstBKT,zhang2018machine,suchsland2018parameter}.
While many conventional, symmetry-breaking, phase transitions can be efficiently captured by ML approaches due to the local nature of the order parameter, topological phase transitions are particularly difficult to ``learn'' \cite{RogerXY}. The difficulty stems from the fact that  topological phase transitions are characterized by the proliferation of non-local, topological defects that are suppressed in the topologically ordered phase \cite{BKTTransition,SubirsReviewTop}.

One ML approach which is commonly used to classify phases of matter, either topological or symmetry-breaking, 
is based on supervised methods. Such methods require prior labeling of the states (i.e., the input data) in well-known regimes, e.g., whether a state belongs to a ferro- or paramagnetic phase \cite{RogerNature}. The supervised step, however, assumes prior knowledge of the underlying phases of the system and, as such, supervised approaches rule out the possibility of learning unknown phases. Contrary to the supervised case, unsupervised methods learn structure from the data itself without the need of prior labeling. As a result, unsupervised ML is particularly useful when the classification (or its mere existence) is not known \textit{a priori}. 

\begin{figure}[b]
  \centering\includegraphics[scale=1.0]{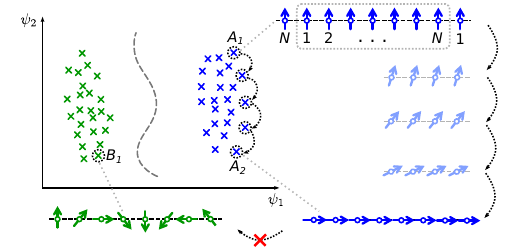}
  \caption{{\bf Topological classification using sample connectivity.} Shown are samples of $N$ classical XY spins, with periodic boundary conditions and winding numbers $\nu = 0,1$, projected on a two-dimensional feature space $\psi_{1,2}$. A diffusion map clusters samples which are connected via continuous deformations, such as $A_1$ and $A_2$, but not $A_1$ and $B_1$.}
  \label{fig:schematics}
\end{figure}

Here we present an unsupervised ML approach that can efficiently classify states according to their topological properties without knowledge of the underlying topological invariant, and can detect topological phase transitions from raw data. The approach is based on dimensional reduction via diffusion maps \cite{Coifman7426,nadler2006diffusion,COIFMAN20065}, which has been used for speaker identification \cite{michalevsky2011speaker} or face recognition \cite{barkan2013fast}, and very naturally implements the notion of homotopy by construction of a diffusion process on the data set (\figref{fig:schematics}). To demonstrate the algorithm, we show that it can efficiently learn the global, topological aspects of the Berezinskii-Kosterlitz-Thouless (BKT) phase transition \cite{BKTTransition,berezinskii,berezinskiiII,KosterlitzII}, which has proven to be notoriously difficult with other ML techniques: as discussed in detail in \refcite{RogerXY}, significant feature engineering, such as using vorticity as input, is required to learn the BKT transition rather than the magnetization of finite-sized samples \cite{zhang2018machine,TrebstBKT,suchsland2018parameter,Wetzel,HuUnsupervised,WangUnsupervised,wang2018machine,cristoforettiUnsupervised}. We also demonstrate that the topological aspects of the crossover from topological order (at zero temperature) to the high-temperature phase in the Ising gauge theory \cite{WegnerIGT,KogutReview,SubirsReviewTop} can be efficiently captured by our proposed scheme.

\sect{Winding number and diffusion maps}To introduce the method, let us consider one of the simplest examples of a topological invariant, the winding number $\nu$ associated with continuous mappings ${\rm S}_1\rightarrow {\rm S}_1$ between two one-spheres: $\theta \in [0,2\pi) \rightarrow \vec{S}(\theta)\in \mathbbm{R}^2$ with $\vec{S}^2=1$ and $\vec{S}(0)=\vec{S}(2\pi)$. The winding number $\nu \in \mathbbm{Z}$ is given by the number of times the unit vector $\vec{S}$ winds around upon traversing ${\rm S}_1$, and is a topological invariant because it cannot be changed by continuous deformations of the mapping. The winding number plays a crucial role in various contexts, such as the classification of topological band insulators \cite{SSHModel} and superconductors \cite{KitaevModel}, or as the topological invariant associated with a non-contractible loop of the quasi-long-range-ordered phase of the XY model on a torus \cite{BKTTransition}. Rather than learning $\nu$ in a supervised way as recently done in \refcite{S1S1MappingSupervised}, here we address the unsupervised version of the problem: classifying different samples of the mapping ${\rm S}_1\rightarrow{\rm S}_1$ without providing the labels $\nu$ for each sample, or the number of distinct $\nu$ which are present in the data set.

To treat the problem numerically, we discretize the mapping $\vec{S}_i = \vec{S}(2\pi i/N)$, $i=1,\dots,N$,  which can be viewed as a one-dimensional (1D) chain of $N$ classical XY spins with periodic boundary conditions (\figref{fig:schematics}). We consider $m$ samples of XY spin configurations, which we denote $x_l =\{\vec{S}^{(l)}_{i} \}$, $l=1,\dots , m$. To define the notion of local similarity between samples $l$ and $l'$, we use a function of the Euclidean norm, for instance the Gaussian kernel
\begin{equation}
K_\epsilon(x_l,x_{l'}) = \exp\left(-\frac{||x_l-x_{l'}||^2}{2 N \epsilon}\right), \label{GeneralKernel}
\end{equation}
with variance controlled by $\epsilon$. In the 1D XY model, the  Euclidean distance between $x_l$ and $x_{l'}$ is given by $||x_l-x_{l'}||^2/(2N) = 1 - \frac{1}{N}\sum_{i=1}^{N}\vec{S}_i^{(l)}\cdot \vec{S}_i^{(l')}$. As a result, \equref{GeneralKernel} quantifies the similarity between two samples by \textit{local} comparison of their degrees of freedom such that $K_\epsilon(x_l,x_{l'})\approx 1$ if $x_{l'}$ is a small deformation of $x_l$.

Importantly, \equref{GeneralKernel} illustrates why unsupervised classification of topologically distinct mappings is a challenging task for conventional ML algorithms, such as Principal Component Analysis (PCA), without \textit{a priori} knowledge of the explicit form of the topological invariant. In particular, two topologically equivalent samples, such as $A_1$ and $A_2$ in \figref{fig:schematics}, can be equally separated from each other in Euclidean space as two topologically distinct samples, such as $A_1$ and $B_1$ in \figref{fig:schematics}, i.e.,~$K_\epsilon(x_{A_1},x_{A_2}) = K_\epsilon(x_{A_1},x_{B_1}) = e^{-1/\epsilon}$. As a result, methods that rely on statistics of the Euclidean distance, such as PCA, are not suitable for topological classification (see examples in \SuplDisc{1 and 2}). 
Rather than learning from {\it local similarity} between individual samples, our goal is to learn from the {\it global connectivity} of the entire sample space. Quantifying connectivity of samples is the essence of diffusion maps, first introduced by Coifman {\it et al}.~\cite{Coifman7426} for manifold learning and dimensional reduction. Here we extend the applicability of this method to perform topological classification of phases of matter. 

A diffusion map is obtained by constructing a diffusion process on the data set $X=\{x_l | l=1,\dots, m \}$. The one-step transition probability $P_{l,l'}$ from sample $l$ to sample $l'$ is defined as
\begin{equation}
P_{l,l'} = \frac{K_{\epsilon}(x_l,x_{l'})}{z_l}, \quad z_l = \sum_{l'=1}^m K_\epsilon(x_l,x_{l'}), \label{DefinitionOfP}
\end{equation}
where $z_l$ has been introduced to account for probability conservation, $\sum_{l'}P_{l,l'}=1$. The value of $z_l$ quantifies the effective coordination number of sample $l$. Crucially, $P_{l,l'}$ and $K_\epsilon (x_l,x_{l'})$ in \equref{GeneralKernel} 
only contain information about the \textit{local} structure of the data. However, the transition probabilities $(P^t)_{l,l'}$ after many $t \in \mathbbm{N}$ diffusion steps experience the \textit{global} structure of the samples. 

\begin{figure}
  \centering\includegraphics[scale=1.2]{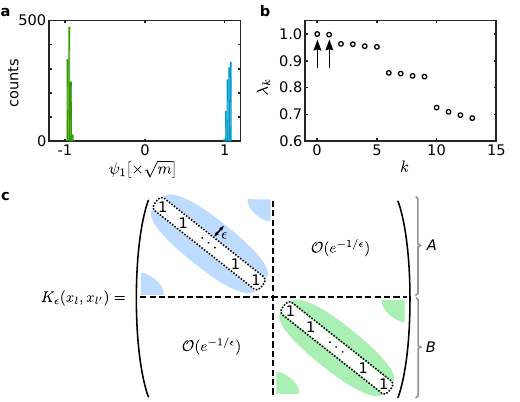}
  \caption{{\bf Detection of topological order in the one-dimensional XY model.} {\bf a}, Clustering of the first component of the diffusion map, $\psi_1$, according to winding number. Shown is a color-coded histogram of $\{(\psi_1)_l|l=1,\ldots,m\}$, with $m$ the number of samples. The color code indicates the (hidden) winding number associated to each bar. {\bf b}, Largest eigenvalues of the transition probability matrix, $P_{l,l'}$ in \equref{DefinitionOfP}. The two-fold degeneracy of the largest eigenvalue (indicated with arrows) signals the presence of two topologically distinct sectors. {\bf c}, Schematics of $K_\epsilon$ in \equref{GeneralKernel} illustrating connectivity between samples for a given $\epsilon$ ($\epsilon$: variance of the Gaussian kernel) and two topological sectors $A$ and $B$. Parameters used in {\bf a} and {\bf b}: $m=2100$, $N=64$, $\sigma_\theta = \pi/5$. 
  }
  \label{fig:histograms}
\end{figure}

More explicitly, the ``diffusion distance'' between samples $l$ and $l'$ after $2t$ times steps can be defined as 
\begin{equation}
D_{t}(l,l') := \sum_{l''} \frac{1}{z_{l''}} \left[ (P^t)_{_{l,l''}} - (P^t)_{_{l',l''}} \right]^2 \geq 0. \label{DefinitionOfD}
\end{equation}
The value of $D_t(l,l')$ is small (large) if there are many (very few) strong paths of length $2t$ connecting the two samples, like in case of $A_1$ and $A_2$ ($A_1$ and $B_1$) in \figref{fig:schematics}. The notion of the diffusion distance motivates the introduction of the \textit{diffusion map}, defined as the mapping
\begin{equation}
 x_l \, \rightarrow \, \vec{\Phi}_l : = \left[ (\psi_1)_l, (\psi_2)_l,\dots,  (\psi_{m-1})_l\right]. \label{DiffusionMap}
\end{equation} 
Here $\psi_k$ are the right eigenvectors of the transition matrix $P \psi_k = \lambda_k \psi_k$, $k=0,1,\dots m-1$, with eigenvalues $\lambda_k \le 1$, and which we order such that $|\lambda_k| \geq |\lambda_{k+1}|$ (the spectrum of $P$ can be efficiently obtained by diagonalizing the similar matrix $A_{l,l'}=P_{l,l'}\sqrt{z_l/z_{l'}}$, which is symmetric). 
The diffusion distance (\ref{DefinitionOfD}) then becomes the weighted Euclidean norm:
\begin{equation}
D_{t}(l,l') = ||\vec{\Phi}_l - \vec{\Phi}_{l'}  ||^2_t = \sum_{k=1}^{m-1} \lambda_k^{2t} [(\psi_k)_l-(\psi_k)_{l'}]^2. \label{RewrittenDiffusionMap}
\end{equation}
Here we dropped the $k=0$ component because it is constant [$(\psi_0)_l = 1/\sqrt{m}$, $\lambda_0=1$] and does not contribute to \equref{RewrittenDiffusionMap}. Because of the weights $\lambda_k^{2t}$ in \equref{RewrittenDiffusionMap}, the long-time $t\rightarrow\infty$ (or global) properties of the samples are encoded in the first few components $(\psi_k)_l$ with largest $\lambda_k$. These few components, which are expected to be similar for samples within the same topological sector in order to make $D_t(l,l')$ small, can be used for dimensional reduction and topological classification, as verified next.

Returning to the mapping ${\rm S}_1\rightarrow{\rm S}_1$, we use randomly generated samples $\vec{S}_i^{(l)} = (\cos\theta_i^{(l)},\sin\theta_i^{(l)})$, with 
\begin{equation}
\theta_i^{(l)} = 2\pi \nu^{(l)} i/N + \delta\theta_i^{(l)} +\bar{\theta}^{(l)}. \label{SimplestWindingData}
\end{equation}
Here $\delta\theta_i^{(l)}$ accounts for spin fluctuations at position $i$, which are taken from a Gaussian distribution with standard deviation $\sigma_\theta$, $\bar{\theta}^{(l)}$ is a random number that represents global rotations, and we use $m =2100$. For now, let us consider having two topological sectors (or winding numbers) in our samples,  $\nu^{(l)}\in \{0,1\}$, which are chosen randomly with equal probability (more general data sets are discussed in the \SuplDisc{2}). 
Figure\,\ref{fig:histograms}a shows the histogram of $(\psi_1)_l$ corresponding to the leading component of the diffusion map. As anticipated, samples with the same $\nu$ have a similar $(\psi_1)_l$ component. Indeed, we find that $(\psi_1)_l \approx +1/\sqrt{m}$ [$(\psi_1)_l \approx -1/\sqrt{m}$] for $\nu=1$ ($\nu=0$) and, as such, $(\psi_1)_l$ clusters samples by winding number. 

For a generic classification problem, the main challenge is that the number $n$ of topological sectors present in the data is not known {\it a priori}. Further, it is unlikely that $(\psi_1)_l$ alone is sufficient for classification. We argue that the answer to both problems is encoded in the eigenvalue spectrum $\lambda_k$. In particular, we find that $n$, and the resulting number of $(\psi_k)_l$ components to be considered for classification, is determined by the degeneracy of the largest eigenvalues $\lambda_k = 1$. For example, as shown in \figref{fig:histograms}b, there are two degenerate eigenvalues $\lambda_{0,1} = 1 $ corresponding to the two winding numbers $\nu = 0,1$, with a clear gap to the subleading $\lambda_{k \geq 2}$. As $\psi_0 = \text{const.}$, the topological classification can be done with $\psi_1$ alone in the simple case above. 

The degeneracy of the largest eigenvalues can be understood from the structure of the matrix $K$ illustrated in \figref{fig:histograms}c upon conveniently reordering the samples by topological sector and increasing value of $\bar{\theta}^{(l)}$ (this does not affect the spectrum of $\lambda_k$ and assumes a small $\sigma_\theta$ such that topological sectors are well-defined). As the two topologically distinct sectors, $A$ and $B$, only couple via exponentially small matrix elements $\sim e^{-1/\epsilon}$, probability conservation in each topological sector separately, $\sum_{l'\in A}P_{l,l'}\approx 1 - {\cal O}(me^{-1/\epsilon})$, ensures the existence of an eigenvalue $\lambda_k\approx 1$ 
up to corrections which can be made exponentially small with increasing sampling size. A derivation of this result can be found in the \SuplDisc{3}. 

\begin{figure}
  \centering\includegraphics[scale=1.0]{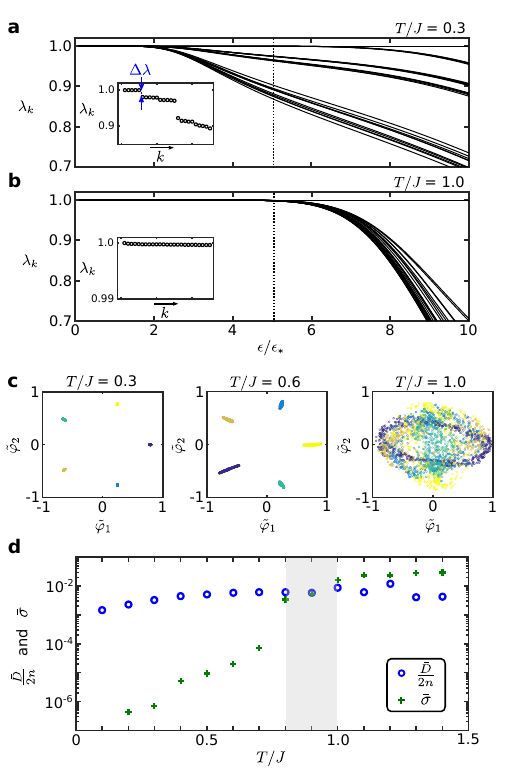}
  \caption{{\bf Detection of topological order in the two-dimensional XY model.} {\bf a,b}, Shown are the largest eigenvalues $\lambda_{k\le 24}$ of $P_{l,l'}$ as a function of $\epsilon$ for ({\bf a}) $T/J = 0.3$ and ({\bf b}) $T/J = 1.0$. The insets show $\lambda_k$ for $\epsilon / \epsilon_* = 5$, with $\epsilon_* = \frac{2\pi}{m_\nu}$ ($m_\nu$: number of samples per sector). {\bf c}, Projection of the 4-dimensional reduced feature space, \equref{ReducedFeatureSpace}, on a two-dimensional plane. The color code refers to the winding number of each sample. Parameters used are described in the main text. {\bf d}, Unsupervised learning of the topological phase transition. Here we exploit the failure of the algorithm to identify the $n$ topological sectors, which leads to a transition temperature of $T_{\rm c}/J = 0.90 \pm 0.1$. $\bar D$ denotes intercluster distance, $\bar\sigma$ is the dispersion of the datapoints between clusters, and $n=5$. The uncertainty of $T_{\rm c}$, indicated with a shaded area, is determined by how much $T_{\rm c}$ changes upon varying $\epsilon$ in the range $\epsilon_*\le\epsilon\le 8\epsilon_*$, see \SuplDisc{6} for details.} 
  \label{fig:2deigv}
\end{figure}

With these observations in mind, we define a reduced ($n-1$)-dimensional ($n\ll m$) feature space $\vec{\varphi}$ given by
\be
x_l \rightarrow \vec{\varphi}_l = [(\psi_1)_l,(\psi_2)_l,\ldots,(\psi_{n-1})_l],
\label{ReducedFeatureSpace}
\ee
where $n$ is the degeneracy of the largest eigenvalue, 
indicating the presence of $n$ topological sectors in sample space. From this low-dimensional feature space, it is straightforward to apply standard clustering algorithms such as $k$-means (see Methods). To illustrate how this general procedure is applied to a specific problem, we now consider the two-dimensional (2D) XY model.

\sect{2D XY model}The configuration energy of the 2D XY model is given by
\be
E(\{\theta_i\}) = -J\sum_{\langle i,j \rangle}  \cos(\theta_i - \theta_j),  \label{XYModelEnergy}
\ee
where $\theta_i$ is the angle of the XY spin at site $i$ on a square lattice, $\langle i,j \rangle$ denotes summation over nearest neighbors, and we use periodic boundary conditions.
This model exhibits the BKT transition \cite{BKTTransition,berezinskii,berezinskiiII,KosterlitzII} at the critical temperature $T_{\rm c}/J \approx 0.89$ \cite{TcBKT}: 
in the topologically ordered phase, $T<T_c$, free vortices are suppressed and two winding numbers, $\nu_x$ and $\nu_y$ along the $x$ and $y$ direction, can be defined to classify globally distinct metastable states \cite{BKTTransition}. The presence of free vortices at $T\geq T_c$ render these invariants ill-defined, which is the hallmark of the BKT transition.

To feed the algorithm, we use spin configurations on a square lattice with $N=32\times 32$ sites, governed by the thermal distribution $\rho(\{\theta_i\}) \propto e^{-{E}(\{\theta_i\})/T}$, and sample over different topological sectors. Specifically, we allow for five possible winding numbers, $(\nu_x,\nu_y) = (0,0)$, $(1,0)$, $(0,1)$, $(-1,0)$, $(0,-1)$ with $m_\nu = 500$ samples each, and use the Metropolis algorithm with local updates in order to thermalize our samples to temperature $T$ (more general data sets with different $\nu_{x,y}$, $m_\nu$ and $N$ are 
discussed in the \SuplDisc{4}). The $ m = 2500$ samples are then shuffled and winding number labels are hidden and stored for later comparison with the output of the diffusion map.  

Figures\,\,\ref{fig:2deigv}a-b show the eigenvalues $\lambda_k$ of $P_{l,l'}$ as function of $\epsilon$, for 
$T<T_{\rm c}$ and $T>T_{\rm c}$, respectively. The degeneracy of the largest eigenvalue (as $\epsilon\rightarrow 0$) signals the number of topological sectors: we find $n = 5$ in \figref{fig:2deigv}a, and $n = 1$ in \figref{fig:2deigv}b. The failure of the algorithm to identify the five different topological sectors at $T/J =1$ is a smoking gun of the topological phase transition.

To illustrate the fidelity of the clustering below $T_c$, we project the low-dimensional feature space $\vec{\varphi}$, Eq.(\ref{ReducedFeatureSpace}), into a two-dimensional plane, $(\tilde{\varphi}_k)_l = \sum_{k'=1}^{n-1}a_{k,k'}(\varphi_{k'})_l$, $k=1,2$, with $a_{k,k'}$ chosen so as to maximize the visibility of clusters. As can be seen in \figref{fig:2deigv}c, the algorithm exhibits 100\% fidelity in the classification for $T<T_{\rm c}$, even for our moderate sampling size and temperatures approaching $T_{\rm c}$ [see extended version of \figref{fig:2deigv}c in the \SuplDisc{5}].

\sect{Learning the topological phase transition}
The failure of the clustering algorithm to identify different topological sectors can be used as diagnostics of the topological phase transition. 
To quantify the success of the clustering algorithm, we use the average intercluster distance ${\bar D}$ and the dispersion ${\bar \sigma}$ of the data points in each cluster, see definitions in Methods. Diffusion maps successfully identifies topological sectors if $\bar\sigma\ll\bar D$, as shown in the first two panels of Fig.~\ref{fig:2deigv}c.  
We define the critical temperature $T_{\rm c}$ as the temperature in which 
$\frac{2\bar\sigma}{\bar D} = \frac{1}{n}$ (i.e.,  
the dispersion of at least one cluster is comparable to half the average distance between clusters). 
As shown in \figref{fig:2deigv}d, using this criterion we find that the topological phase transition occurs at $T/J = 0.90\pm 0.1$, which is in agreement with the known value of $T_{\rm c}/J$ for our system size \cite{RogerXY}. 

\begin{figure}[t]
  \centering\includegraphics[scale=1.2]{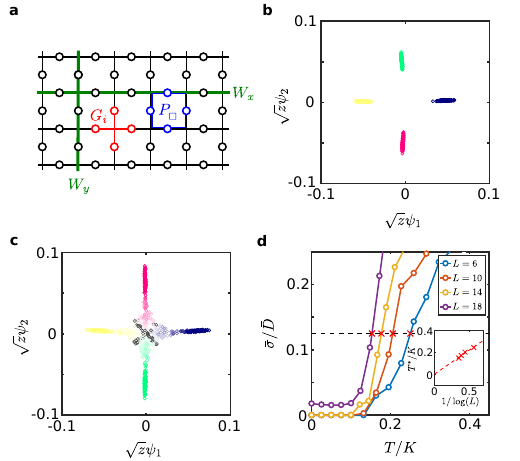}
  \caption{{\bf Detection of topological order in the Ising gauge theory.} 
  {\bf a}, Illustration of the Ising gauge theory showing 
  the Ising degrees of freedom (empty circles), a plaquette operator $P_{_\square}$, a  
  generator $G_i$ of the gauge transformations, and the Wegner-Wilson loops $W_{x,y}$.
  {\bf b,c}, The two leading components of the diffusion map are shown for a system with $N=2\times 14\times 14$ Ising degrees of freedom ($m=800$ samples, $\epsilon=0.02$) for ({\bf b}) $T/K=0.05$ and ({\bf c}) $T/K=0.18$, respectively. 
  The color indicates the topological sector which the respective sample is most closely associated to and its intensity is a measure for how strong the association with the given sector is based on the Wegner-Wilson loops (see Methods).
  The states shown in black cannot be uniquely associated with a single topological sector. {\bf d}, The ratio of within-cluster fluctuation $\bar{\sigma}$ and inter-cluster distance $\bar{D}$ (see Methods) is shown as a function of temperature for different system sizes. Here we use $m=500$ samples and $\epsilon=0.015$. The extracted crossover temperature $T^*$ (defined by $\bar{\sigma}/\bar{D}$ crossing the dashed line) shows the expected logarithmic scaling with system size that extrapolates to zero for $L\rightarrow \infty$ (inset).}
  \label{IsingGaugeTheory}
\end{figure}

\sect{Ising gauge theory} To demonstrate the success of the approach on other models with topological order, we  
apply diffusion maps on the Ising gauge theory. Its configuration energy reads as
\begin{equation}
    E[\{\sigma_{b}\}] = -K \sum_\square P_{_\square}, \quad P_{_\square} \equiv \prod_{b \in \square} \sigma_{b}, \label{IsingGaugeTheoryEn}
\end{equation}
where the Ising spins $\sigma_{b}=\pm 1$ are located on the bonds $b$ of a $L\times L$ square lattice, the sum involves all elementary square plaquettes $\square$, and the product is over all bonds of the plaquette  (\figref{IsingGaugeTheory}a). This model exhibits a local symmetry under a simultaneous flip, $G_i$, of the four spins with a common neighboring vertex $i$ of the square lattice and is, hence, referred to as a gauge theory. The (gauge-invariant) Wegner-Wilson loops $W_{x,y} \in \mathbbm{Z}_2$, defined as the product of $\sigma_{b}$ along non-contractible loops on the torus (\figref{IsingGaugeTheory}a), play the analogous role of $\nu_{x,y}$ in the XY model. In the topologically ordered phase, there are four sectors corresponding to the four different configurations $(W_{x},W_{y})=(\pm 1, \pm 1)$ which are clearly identified by our diffusion map approach (\figref{IsingGaugeTheory}b). To obtain these results, we have used a kernel $K_{l,l'}$ measuring the number of physical spin-flips, i.e., flips of $\sigma_b$ modulo gauge transformations, between two samples $l$ and $l'$: $K_\epsilon(l,l') = \exp[-(1-f_{l,l'})/(N \epsilon)]$ where $f_{l,l'} = \max_{G=\{G_i\}} \sum_{b} (G\sigma^{(l)})^{\phantom{(l)}}_b \hspace{-0.4em} \cdot \sigma_b^{(l')}$ and $(G\sigma^{(l)})^{\phantom{(l)}}_b\hspace{-0.6em}$  denotes the Ising degree of freedom in sample $l$ on bond $b$ after performing the gauge transformation $G$.
The maximization over the exponentially large manifold of possible $G$ can be very efficiently performed stochastically (see Methods) and illustrates the advantage of kernel techniques, such as diffusion maps, for studying gauge theories.

With increasing temperature, different topological sectors become connected via states with strings of flipped Ising spins, the low-energy excitations of the Ising gauge theory. In this case, topological order is lost, which is clearly manifested in the output of 
diffusion maps (\figref{IsingGaugeTheory}c). Interestingly, the output of diffusion maps 
arranges the states according to how they are connected to the different topological sectors, as demonstrated by computing  
the Wegner-Wilson loops (color and intensity of dots in \figref{IsingGaugeTheory}c) for each sample. The connection between the output of diffusion maps and the Wilson-Wegner loop operators is a clearcut evidence that diffusion maps learns the global aspects of the topological phase 
rather than, for instance, local constraints \cite{RogerNature}.

Note that the Ising gauge theory does not exhibit a finite-temperature phase transition and, hence, we can only expect to see a crossover temperature $T^*$ to a non-topological phase in a finite system. Using the same criterion of cluster visibility $\bar{\sigma}/\bar{D}$ as in the 2D XY case, we see from \figref{IsingGaugeTheory}d that the extracted values of $T^*$ based on diffusion maps shows the expected logarithmic scaling with system size $L$ that extrapolates to zero as $L\rightarrow \infty$.


\begin{figure}[t]
  \centering\includegraphics[scale=1.3]{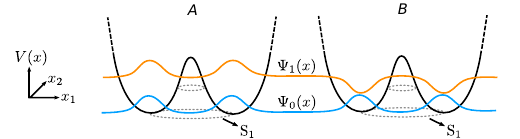}
  \caption{{\bf Illustration of the diffusion map in the continuum limit.} Shown are the degenerate ground states $\Psi_0$ and $\Psi_1$ in the case of two topologically distinct sectors, $A$ and $B$, which correspond to the minima of the potential $V(x)$, see Eq.~(\ref{GeneralHamiltonian}).}
  \label{DualProblem}
\end{figure}
  
\sect{Diffusion maps in the continuum limit} At low temperatures, the output of the diffusion map, $(\psi_k)_l$ and $\lambda_k$, can be understood in terms of the eigenstates $\Psi_k$ and eigenvalues $E_k$ of a many-body quantum well Hamiltonian. Let us first focus on a single topological sector, and take $m \rightarrow \infty$. In this limit, the summations over sample space can be interpreted as Monte Carlo approximations of integrals, e.g., $z_l \sim z(x_l) = \int_{\cal C}\diff x' K(x_l,x')\rho(x') $ in  \equref{DefinitionOfP}, where $x'$ is integrated over configuration space ${\cal C}$, with a probability distribution $\rho(x') \propto e^{-E(x')/T}$ and energy $E(x')$. It was shown in \refcite{COIFMAN20065} that, for small $\epsilon$, the eigenvectors and eigenvalues of $P_{l,l'}$ can be written as  $ (\psi_k)_l \sim \Psi_k(x_l)/\rho(x_l)$ and $\lambda_k \sim 1 - \epsilon E_k$, where $\Psi_k$ and $E_k$ are the eigenstates and energies of 
\begin{equation}
\hat{\mathcal{H}} = \frac{\hat{\boldsymbol p}^2}{2} + V(x), \quad V(x) =  \frac{||\nabla_x E||^2}{2T^2} - \frac{\Delta_x E}{2T}. \label{GeneralHamiltonian}
\end{equation}
Here $\hat{\boldsymbol p} = -i {\nabla}_x$ is the momentum operator, ${\nabla}_x$ is the gradient on $\mathcal{C}$, and ${\Delta}_x = {\nabla}_x \cdot {\nabla}_x$ is the Laplacian. For instance, in the XY model in \equref{XYModelEnergy}, we have $\mathcal{C}=\textrm{S}_1^N$ and $\hat{\boldsymbol p} = -i(\partial_{\theta_1},\ldots,\partial_{\theta_N})$. As illustrated in \figref{DualProblem}, $V(x)$ resembles a quantum well with the minima occurring on a submanifold of ${\cal C}$ where $E$ is minimized. Crucial for topological classification is the existence of the ``zero mode'' $\Psi_0(x) = \rho(x)$ of $\hat{\mathcal{H}}$, i.e., with $E_0 =0$ (or $\lambda_0 = 1$).

In the presence of, say, two topological sectors, $A$ and $B$, each of which provides a zero mode, there will be a two-fold degeneracy, $E_{0,1} = 0$ ($\lambda_{0,1} =1 $). In this case, the quantum system resembles a double quantum well with high potential barriers and exponentially small tunneling coupling (\figref{DualProblem}). The two-fold degenerate ground state will be formed by an even ($k=0$) and an odd ($k=1$) eigenstate with exponentially small splitting (see \SuplDisc{2}): $\Psi_0(x) = \rho(x)$, $\forall x$, and $\Psi_1(x) \approx \rho(x)$ if $x\in A$,  $\Psi_1(x) \approx -\rho(x)$ if $x\in B$.  
As a result, the sign of the antisymmetric state $\Psi_1(x)$ encodes the topological sector of $x$, as previously shown in \figref{fig:histograms}a. Extending this argument to $n$ topological sectors with $n$ zero modes is straightforward.

To summarize, we have shown that diffusion maps can be used for topological classification and to learn topological phase transitions in an unsupervised fashion. The success of this approach on learning the BKT transition and detecting topological order in the Ising gauge theory suggests that diffusion maps can complement other versatile machine learning methods, such as those based on neural networks, in order to perform fully unsupervised studies of topological states of matter.

\vspace{1em}

\begin{center}
    \textbf{Methods}
\end{center}
 
\sect{Clustering in the reduced feature space}After dimensional reduction using the diffusion map, we use the $k$-means algorithm, which is a standard clustering algorithm for unsupervised machine learning \cite{NielsensBook,GoodfellowBook}, to classify the samples. In our case, each sample $x_l$ has an associated $(n-1)$-dimensional feature vector $\vec{\varphi}_l = [(\psi_1)_l,\ldots,(\psi_{n-1})_l]$, see \equref{ReducedFeatureSpace} in the main text, formed by the leading eigenvectors of the transition matrix $P_{l,l'}$. As already discussed in the main text, the zeroth eigenvector, $(\psi_0)_l$, has the same components for all samples and can, hence, be discarded for labeling purposes. 

The objective of $k$-means is to partition the $m$ samples into $k$ subsets $S_j$, $j = 1,\ldots,k$, in order to minimize the function
\be
{\cal F} =  \sum_{j = 1}^{k}\sum_{l\in S_j}||\vec{\varphi}_l - {\boldsymbol\mu}_j||^2,
\label{eq:functional}
\ee
where $\vec{\mu}_j$ is the center of mass of subset $S_j$. Once \equref{eq:functional} is minimized, each sample $l\in S_{j}$ is given the label $j$. As follows from our discussion of the main text, we use $k=n$ as the number of clusters to classify samples according to their topological properties. 

As a quantitative measure of the visibility of clusters, which we use to extract a critical temperature for the topological phase transitions, we take the ratio $\bar{\sigma}/\bar{D}$ of the average within-cluster standard deviation,
\begin{equation}
    \bar{\sigma} = \frac{1}{n} \sum_{j = 1}^{n} \sqrt{ \frac{1}{|S_j|} \sum_{l\in S_j}||\vec{\varphi}_l - {\boldsymbol\mu}_j||^2},
\end{equation}
where $|S_j|$ is the number of samples in the cluster $j$, and the average inter-cluster distance,
\begin{equation}
    \bar{D} = \frac{1}{n(n-1)}\sum_{j,j'=1}^n ||{\boldsymbol\mu}_j - {\boldsymbol\mu}_{j'}||.
\end{equation}

\sect{Samples for the Ising gauge theory} In a fixed gauge, the spectrum of the Ising gauge theory in \equref{IsingGaugeTheoryEn} is given by four ground states, distinguished by $W_{x,y}$, and pairs of vison excitations corresponding to the endpoints of ``strings'' along which Ising spins are flipped \cite{SubirsReviewTop}. Instead of performing Monte Carlo simulations, we have generated data by randomly sampling from all four ground states, adding (with the correct thermal probability) visons at random positions, and finally performing random gauge transformations. This has the advantage of yielding statistically independent configurations. For simplicity, we have focused on the subspace of at most one pair of visons.

As discussed in the main text, we have to maximize $\sum_{b} (G\sigma^{(l)})^{\phantom{(l)}}_b \hspace{-0.4em} \cdot \sigma_b^{(l')}$ over the gauge transformations $G=\{G_i\}$ in order to compute the kernel $K_{l,l'}$. As the set of possible gauge transformations is exponentially large ($2^{L^2} \approx 10^{30}$ for $L=10$), exact maximization is not possible. Instead, we randomly choose one site $i=i_0$ and apply $G_{i_0}$. We keep the gauge transformation if $\sum_{b} (G\sigma^{(l)})^{\phantom{(l)}}_b \hspace{-0.4em} \cdot \sigma_b^{(l')}$ has not decreased when $G\rightarrow G_{i_0} G $ and repeat $N_G$ times. For instance, for $L=10$, $N_G=2000$ is sufficient. In \figref{IsingGaugeTheory}, we used $N_G=1000$ for $L=6$, $N_G=2000$ for $L=10$, $N_G=3000$ for $L=14$, and $N_G=7000$ for $L=18$.

\sect{Wegner-Wilson loops} To provide a direct interpretation of what the diffusion map has learnt from the data, we have computed the values of the Wegner-Wilson loops along all different non-contractible straight lines, 
\begin{equation}
    W^{(l)}_x(i_y) = \prod_{i_x=1}^L \sigma^{(l)}_{(i,i+\vec{e}_x)}, \quad W^{(l)}_y(i_x) = \prod_{i_y=1}^L \sigma^{(l)}_{(i,i+\vec{e}_y)}, \nonumber
\end{equation}
for each sample $l$. We associate sample $l$ with a topological sector and, hence, a color in \figref{IsingGaugeTheory}(b) and (c), by majority vote, i.e., use $\left(\sign( \overline{W}^{(l)}_x),\sign( \overline{W}^{(l)}_y)\right)$ as topological indices where $\overline{W}^{(l)}_{x,y}=\sum_{i_{y,x}} W^{(l)}_{x,y}(i_{y,x})/L$ is the average Wegner-Wilson loop. If either of $\overline{W}^{(l)}_{x}$ and $\overline{W}^{(l)}_{y}$ vanishes, the corresponding state is equally strong associated with (at least) two different topological sectors and is labeled in black in the figure. As a measure for how strongly a given state $l$ is associated with the topological sector it is closest to, we count the number of times $W_{x,y}^{(l)}(i_{y,x})\neq \sign(\overline{W}^{(l)}_{x,y})$ for $i_{y,x} = 1, \dots, L$, which will be denoted by $\Delta W^{(l)}$; the smaller $\Delta W^{(l)}$, the stronger the association with the topological sector and the higher the intensity of the respective dot in \figref{IsingGaugeTheory}b and c. 

We have also verified that the average of $\Delta W^{(l)}$ over samples of a system at a given temperature $T$, can be used as a criterion for the crossover temperature and that the resulting $T^*$ also shows the logarithmic scaling behavior as seen in the diffusion map approach.

\sect{Code availability} Code used to generate the results presented in this work can be found in the online version of the paper.

\sect{Data availability}The data that support the plots within this paper and other findings of this study are available from the corresponding author upon reasonable request.


\begin{thebibliography}{49}
\expandafter\ifx\csname url\endcsname\relax
  \def\url#1{\texttt{#1}}\fi
\expandafter\ifx\csname urlprefix\endcsname\relax\def\urlprefix{URL }\fi
\providecommand{\bibinfo}[2]{#2}
\providecommand{\eprint}[2][]{\url{#2}}

\bibitem{NielsensBook}
\bibinfo{author}{Nielsen, M.~A.}
\newblock \emph{\bibinfo{title}{Neural Networks and Deep Learning}}
  (\bibinfo{publisher}{Determination Press}, \bibinfo{year}{2015}).

\bibitem{GoodfellowBook}
\bibinfo{author}{Goodfellow, I.}, \bibinfo{author}{Bengio, Y.} \&
  \bibinfo{author}{Courville, A.}
\newblock \emph{\bibinfo{title}{Deep Learning}} (\bibinfo{publisher}{MIT
  Press}, \bibinfo{year}{2016}).
\newblock \bibinfo{note}{\url{http://www.deeplearningbook.org}}.

\bibitem{mehta2018Review}
\bibinfo{author}{Mehta, P.} \emph{et~al.}
\newblock \bibinfo{title}{A high-bias, low-variance introduction to machine
  learning for physicists}.
\newblock \emph{\bibinfo{journal}{arXiv preprint arXiv:1803.08823}}
  (\bibinfo{year}{2018}).

\bibitem{ConnectionToRG}
\bibinfo{author}{Mehta, P.} \& \bibinfo{author}{Schwab, D.~J.}
\newblock \bibinfo{title}{An exact mapping between the variational
  renormalization group and deep learning}.
\newblock \emph{\bibinfo{journal}{arXiv preprint arXiv:1410.3831}}
  (\bibinfo{year}{2014}).

\bibitem{Carleo602}
\bibinfo{author}{Carleo, G.} \& \bibinfo{author}{Troyer, M.}
\newblock \bibinfo{title}{Solving the quantum many-body problem with artificial
  neural networks}.
\newblock \emph{\bibinfo{journal}{Science}} \textbf{\bibinfo{volume}{355}},
  \bibinfo{pages}{602--606} (\bibinfo{year}{2017}).

\bibitem{Sarma}
\bibinfo{author}{{Deng}, D.-L.}, \bibinfo{author}{{Li}, X.} \&
  \bibinfo{author}{{Das Sarma}, S.}
\newblock \bibinfo{title}{{Machine learning topological states}}.
\newblock \emph{\bibinfo{journal}{Phys. Rev. B}} \textbf{\bibinfo{volume}{96}},
  \bibinfo{pages}{195145} (\bibinfo{year}{2017}).

\bibitem{FuSelfLearning}
\bibinfo{author}{Liu, J.}, \bibinfo{author}{Qi, Y.}, \bibinfo{author}{Meng,
  Z.~Y.} \& \bibinfo{author}{Fu, L.}
\newblock \bibinfo{title}{Self-learning {Monte Carlo} method}.
\newblock \emph{\bibinfo{journal}{Phys. Rev. B}} \textbf{\bibinfo{volume}{95}},
  \bibinfo{pages}{041101} (\bibinfo{year}{2017}).

\bibitem{QuantumStateTomography}
\bibinfo{author}{Torlai, G.} \emph{et~al.}
\newblock \bibinfo{title}{Neural-network quantum state tomography}.
\newblock \emph{\bibinfo{journal}{Nature Physics}}  (\bibinfo{year}{2018}).

\bibitem{YZYouGeometry}
\bibinfo{author}{You, Y.-Z.}, \bibinfo{author}{Yang, Z.} \&
  \bibinfo{author}{Qi, X.-L.}
\newblock \bibinfo{title}{Machine learning spatial geometry from entanglement
  features}.
\newblock \emph{\bibinfo{journal}{Phys. Rev. B}} \textbf{\bibinfo{volume}{97}},
  \bibinfo{pages}{045153} (\bibinfo{year}{2018}).

\bibitem{carleo2018constructing}
\bibinfo{author}{{Carleo}, G.}, \bibinfo{author}{{Nomura}, Y.} \&
  \bibinfo{author}{{Imada}, M.}
\newblock \bibinfo{title}{{Constructing exact representations of quantum
  many-body systems with deep neural networks}}.
\newblock \emph{\bibinfo{journal}{Nature Communications}}
  \textbf{\bibinfo{volume}{9}}, \bibinfo{pages}{5322} (\bibinfo{year}{2018}).

\bibitem{InteractionDistance}
\bibinfo{author}{Spillard, S.}, \bibinfo{author}{Turner, C.~J.} \&
  \bibinfo{author}{Meichanetzidis, K.}
\newblock \bibinfo{title}{Machine learning entanglement freedom}.
\newblock \emph{\bibinfo{journal}{International Journal of Quantum
  Information}} \textbf{\bibinfo{volume}{16}}, \bibinfo{pages}{1840002}
  (\bibinfo{year}{2018}).

\bibitem{RogerNature}
\bibinfo{author}{Carrasquilla, J.} \& \bibinfo{author}{Melko, R.~G.}
\newblock \bibinfo{title}{Machine learning phases of matter}.
\newblock \emph{\bibinfo{journal}{Nature Physics}}
  \textbf{\bibinfo{volume}{13}}, \bibinfo{pages}{431} (\bibinfo{year}{2017}).

\bibitem{MelkoFermions}
\bibinfo{author}{Ch'ng, K.}, \bibinfo{author}{Carrasquilla, J.},
  \bibinfo{author}{Melko, R.~G.} \& \bibinfo{author}{Khatami, E.}
\newblock \bibinfo{title}{Machine learning phases of strongly correlated
  fermions}.
\newblock \emph{\bibinfo{journal}{Phys. Rev. X}} \textbf{\bibinfo{volume}{7}},
  \bibinfo{pages}{031038} (\bibinfo{year}{2017}).

\bibitem{PhysRevLett.118.216401}
\bibinfo{author}{Zhang, Y.} \& \bibinfo{author}{Kim, E.-A.}
\newblock \bibinfo{title}{Quantum loop topography for machine learning}.
\newblock \emph{\bibinfo{journal}{Phys. Rev. Lett.}}
  \textbf{\bibinfo{volume}{118}}, \bibinfo{pages}{216401}
  (\bibinfo{year}{2017}).

\bibitem{QuantumLoopTopography2}
\bibinfo{author}{Zhang, Y.}, \bibinfo{author}{Melko, R.~G.} \&
  \bibinfo{author}{Kim, E.-A.}
\newblock \bibinfo{title}{Machine learning {${\mathbb{Z}}_{2}$} quantum spin
  liquids with quasiparticle statistics}.
\newblock \emph{\bibinfo{journal}{Phys. Rev. B}} \textbf{\bibinfo{volume}{96}},
  \bibinfo{pages}{245119} (\bibinfo{year}{2017}).

\bibitem{Confusion}
\bibinfo{author}{van Nieuwenburg, E. P.~L.}, \bibinfo{author}{Liu, Y.-H.} \&
  \bibinfo{author}{Huber, S.~D.}
\newblock \bibinfo{title}{Learning phase transitions by confusion}.
\newblock \emph{\bibinfo{journal}{Nature Physics}}
  \textbf{\bibinfo{volume}{13}}, \bibinfo{pages}{435--439}
  (\bibinfo{year}{2017}).

\bibitem{Rand2DSystem}
\bibinfo{author}{Ohtsuki, T.} \& \bibinfo{author}{Ohtsuki, T.}
\newblock \bibinfo{title}{Deep learning the quantum phase transitions in random
  two-dimensional electron systems}.
\newblock \emph{\bibinfo{journal}{Journal of the Physical Society of Japan}}
  \textbf{\bibinfo{volume}{85}}, \bibinfo{pages}{123706}
  (\bibinfo{year}{2016}).

\bibitem{RandomSystems}
\bibinfo{author}{Ohtsuki, T.} \& \bibinfo{author}{Ohtsuki, T.}
\newblock \bibinfo{title}{Deep learning the quantum phase transitions in random
  electron systems: Applications to three dimensions}.
\newblock \emph{\bibinfo{journal}{Journal of the Physical Society of Japan}}
  \textbf{\bibinfo{volume}{86}}, \bibinfo{pages}{044708}
  (\bibinfo{year}{2017}).

\bibitem{yoshiokaDisorder}
\bibinfo{author}{{Yoshioka}, N.}, \bibinfo{author}{{Akagi}, Y.} \&
  \bibinfo{author}{{Katsura}, H.}
\newblock \bibinfo{title}{{Learning disordered topological phases by
  statistical recovery of symmetry}}.
\newblock \emph{\bibinfo{journal}{Phys. Rev. B}} \textbf{\bibinfo{volume}{97}},
  \bibinfo{pages}{205110} (\bibinfo{year}{2018}).

\bibitem{S1S1MappingSupervised}
\bibinfo{author}{Zhang, P.}, \bibinfo{author}{Shen, H.} \&
  \bibinfo{author}{Zhai, H.}
\newblock \bibinfo{title}{Machine learning topological invariants with neural
  networks}.
\newblock \emph{\bibinfo{journal}{Phys. Rev. Lett.}}
  \textbf{\bibinfo{volume}{120}}, \bibinfo{pages}{066401}
  (\bibinfo{year}{2018}).

\bibitem{RealSpaceTopInvs}
\bibinfo{author}{Carvalho, D.}, \bibinfo{author}{Garc\'{\i}a-Mart\'{\i}nez,
  N.~A.}, \bibinfo{author}{Lado, J.~L.} \&
  \bibinfo{author}{Fern\'andez-Rossier, J.}
\newblock \bibinfo{title}{Real-space mapping of topological invariants using
  artificial neural networks}.
\newblock \emph{\bibinfo{journal}{Phys. Rev. B}} \textbf{\bibinfo{volume}{97}},
  \bibinfo{pages}{115453} (\bibinfo{year}{2018}).

\bibitem{AdversarialNNs}
\bibinfo{author}{Huembeli, P.}, \bibinfo{author}{Dauphin, A.} \&
  \bibinfo{author}{Wittek, P.}
\newblock \bibinfo{title}{Identifying quantum phase transitions with
  adversarial neural networks}.
\newblock \emph{\bibinfo{journal}{Phys. Rev. B}} \textbf{\bibinfo{volume}{97}},
  \bibinfo{pages}{134109} (\bibinfo{year}{2018}).

\bibitem{iakovlev2018supervised}
\bibinfo{author}{Iakovlev, I.~A.}, \bibinfo{author}{Sotnikov, O.~M.} \&
  \bibinfo{author}{Mazurenko, V.~V.}
\newblock \bibinfo{title}{Supervised learning approach for recognizing magnetic
  skyrmion phases}.
\newblock \emph{\bibinfo{journal}{Phys. Rev. B}} \textbf{\bibinfo{volume}{98}},
  \bibinfo{pages}{174411} (\bibinfo{year}{2018}).

\bibitem{vargas2018extrapolating}
\bibinfo{author}{Vargas-Hern\'andez, R.~A.}, \bibinfo{author}{Sous, J.},
  \bibinfo{author}{Berciu, M.} \& \bibinfo{author}{Krems, R.~V.}
\newblock \bibinfo{title}{Extrapolating quantum observables with machine
  learning: Inferring multiple phase transitions from properties of a single
  phase}.
\newblock \emph{\bibinfo{journal}{Phys. Rev. Lett.}}
  \textbf{\bibinfo{volume}{121}}, \bibinfo{pages}{255702}
  (\bibinfo{year}{2018}).

\bibitem{RogerXY}
\bibinfo{author}{Beach, M. J.~S.}, \bibinfo{author}{Golubeva, A.} \&
  \bibinfo{author}{Melko, R.~G.}
\newblock \bibinfo{title}{Machine learning vortices at the
  {Kosterlitz-Thouless} transition}.
\newblock \emph{\bibinfo{journal}{Phys. Rev. B}} \textbf{\bibinfo{volume}{97}},
  \bibinfo{pages}{045207} (\bibinfo{year}{2018}).

\bibitem{wang2018machine}
\bibinfo{author}{Wang, C.} \& \bibinfo{author}{Zhai, H.}
\newblock \bibinfo{title}{Machine learning of frustrated classical spin models
  (ii): Kernel principal component analysis}.
\newblock \emph{\bibinfo{journal}{Frontiers of Physics}}
  \textbf{\bibinfo{volume}{13}}, \bibinfo{pages}{130507}
  (\bibinfo{year}{2018}).

\bibitem{HuUnsupervised}
\bibinfo{author}{Hu, W.}, \bibinfo{author}{Singh, R. R.~P.} \&
  \bibinfo{author}{Scalettar, R.~T.}
\newblock \bibinfo{title}{Discovering phases, phase transitions, and crossovers
  through unsupervised machine learning: A critical examination}.
\newblock \emph{\bibinfo{journal}{Phys. Rev. E}} \textbf{\bibinfo{volume}{95}},
  \bibinfo{pages}{062122} (\bibinfo{year}{2017}).

\bibitem{Wetzel}
\bibinfo{author}{Wetzel, S.~J.}
\newblock \bibinfo{title}{Unsupervised learning of phase transitions: From
  principal component analysis to variational autoencoders}.
\newblock \emph{\bibinfo{journal}{Phys. Rev. E}} \textbf{\bibinfo{volume}{96}},
  \bibinfo{pages}{022140} (\bibinfo{year}{2017}).

\bibitem{WangUnsupervised}
\bibinfo{author}{Wang, C.} \& \bibinfo{author}{Zhai, H.}
\newblock \bibinfo{title}{Machine learning of frustrated classical spin models.
  i. principal component analysis}.
\newblock \emph{\bibinfo{journal}{Phys. Rev. B}} \textbf{\bibinfo{volume}{96}},
  \bibinfo{pages}{144432} (\bibinfo{year}{2017}).

\bibitem{cristoforettiUnsupervised}
\bibinfo{author}{Cristoforetti, M.}, \bibinfo{author}{Jurman, G.},
  \bibinfo{author}{Nardelli, A.~I.} \& \bibinfo{author}{Furlanello, C.}
\newblock \bibinfo{title}{Towards meaningful physics from generative models}.
\newblock \emph{\bibinfo{journal}{arXiv preprint arXiv:1705.09524}}
  (\bibinfo{year}{2017}).

\bibitem{TrebstBKT}
\bibinfo{author}{Broecker, P.}, \bibinfo{author}{Assaad, F.~F.} \&
  \bibinfo{author}{Trebst, S.}
\newblock \bibinfo{title}{Quantum phase recognition via unsupervised machine
  learning}.
\newblock \emph{\bibinfo{journal}{arXiv preprint arXiv:1707.00663}}
  (\bibinfo{year}{2017}).

\bibitem{zhang2018machine}
\bibinfo{author}{Zhang, W.}, \bibinfo{author}{Liu, J.} \& \bibinfo{author}{Wei,
  T.-C.}
\newblock \bibinfo{title}{Machine learning of phase transitions in the
  percolation and xy models}.
\newblock \emph{\bibinfo{journal}{arXiv preprint arXiv:1804.02709}}
  (\bibinfo{year}{2018}).

\bibitem{suchsland2018parameter}
\bibinfo{author}{{Suchsland}, P.} \& \bibinfo{author}{{Wessel}, S.}
\newblock \bibinfo{title}{{Parameter diagnostics of phases and phase transition
  learning by neural networks}}.
\newblock \emph{\bibinfo{journal}{Physical Review B}}
  \textbf{\bibinfo{volume}{97}}, \bibinfo{pages}{174435}
  (\bibinfo{year}{2018}).

\bibitem{BKTTransition}
\bibinfo{author}{Kosterlitz, J.~M.} \& \bibinfo{author}{Thouless, D.~J.}
\newblock \bibinfo{title}{Ordering, metastability and phase transitions in
  two-dimensional systems}.
\newblock \emph{\bibinfo{journal}{Journal of Physics C: Solid State Physics}}
  \textbf{\bibinfo{volume}{6}}, \bibinfo{pages}{1181} (\bibinfo{year}{1973}).

\bibitem{SubirsReviewTop}
\bibinfo{author}{{Sachdev}, S.}
\newblock \bibinfo{title}{{Topological order, emergent gauge fields, and Fermi
  surface reconstruction}}.
\newblock \emph{\bibinfo{journal}{Reports on Progress in Physics}}
  \textbf{\bibinfo{volume}{82}}, \bibinfo{pages}{014001}
  (\bibinfo{year}{2019}).

\bibitem{Coifman7426}
\bibinfo{author}{Coifman, R.~R.} \emph{et~al.}
\newblock \bibinfo{title}{Geometric diffusions as a tool for harmonic analysis
  and structure definition of data: Diffusion maps}.
\newblock \emph{\bibinfo{journal}{Proceedings of the National Academy of
  Sciences}} \textbf{\bibinfo{volume}{102}}, \bibinfo{pages}{7426--7431}
  (\bibinfo{year}{2005}).

\bibitem{nadler2006diffusion}
\bibinfo{author}{Nadler, B.}, \bibinfo{author}{Lafon, S.},
  \bibinfo{author}{Kevrekidis, I.} \& \bibinfo{author}{Coifman, R.~R.}
\newblock \bibinfo{title}{Diffusion maps, spectral clustering and
  eigenfunctions of {Fokker-Planck} operators}.
\newblock In \emph{\bibinfo{booktitle}{Advances in neural information
  processing systems}}, \bibinfo{pages}{955--962} (\bibinfo{year}{2006}).

\bibitem{COIFMAN20065}
\bibinfo{author}{Coifman, R.~R.} \& \bibinfo{author}{Lafon, S.}
\newblock \bibinfo{title}{Diffusion maps}.
\newblock \emph{\bibinfo{journal}{Applied and Computational Harmonic Analysis}}
  \textbf{\bibinfo{volume}{21}}, \bibinfo{pages}{5 -- 30}
  (\bibinfo{year}{2006}).
\newblock \bibinfo{note}{Special Issue: Diffusion Maps and Wavelets}.

\bibitem{michalevsky2011speaker}
\bibinfo{author}{Michalevsky, Y.}, \bibinfo{author}{Talmon, R.} \&
  \bibinfo{author}{Cohen, I.}
\newblock \bibinfo{title}{Speaker identification using diffusion maps}.
\newblock In \emph{\bibinfo{booktitle}{Signal Processing Conference, 2011 19th
  European}}, \bibinfo{pages}{1299--1302} (\bibinfo{organization}{IEEE},
  \bibinfo{year}{2011}).

\bibitem{barkan2013fast}
\bibinfo{author}{Barkan, O.}, \bibinfo{author}{Weill, J.},
  \bibinfo{author}{Wolf, L.} \& \bibinfo{author}{Aronowitz, H.}
\newblock \bibinfo{title}{Fast high dimensional vector multiplication face
  recognition}.
\newblock In \emph{\bibinfo{booktitle}{Proceedings of the IEEE International
  Conference on Computer Vision}}, \bibinfo{pages}{1960--1967}
  (\bibinfo{year}{2013}).

\bibitem{berezinskii}
\bibinfo{author}{Berezinskii, V.}
\newblock \bibinfo{title}{Destruction of long-range order in one-dimensional
  and two-dimensional systems having a continuous symmetry group i. classical
  systems}.
\newblock \emph{\bibinfo{journal}{Sov. Phys. JETP}}
  \textbf{\bibinfo{volume}{32}}, \bibinfo{pages}{493--500}
  (\bibinfo{year}{1971}).

\bibitem{berezinskiiII}
\bibinfo{author}{Berezinskii, V.}
\newblock \bibinfo{title}{Destruction of long-range order in one-dimensional
  and two-dimensional systems possessing a continuous symmetry group. ii.
  quantum systems}.
\newblock \emph{\bibinfo{journal}{Soviet Journal of Experimental and
  Theoretical Physics}} \textbf{\bibinfo{volume}{34}}, \bibinfo{pages}{610}
  (\bibinfo{year}{1972}).

\bibitem{KosterlitzII}
\bibinfo{author}{Kosterlitz, J.~M.}
\newblock \bibinfo{title}{The critical properties of the two-dimensional xy
  model}.
\newblock \emph{\bibinfo{journal}{Journal of Physics C: Solid State Physics}}
  \textbf{\bibinfo{volume}{7}}, \bibinfo{pages}{1046} (\bibinfo{year}{1974}).

\bibitem{WegnerIGT}
\bibinfo{author}{Wegner, F.~J.}
\newblock \bibinfo{title}{Duality in generalized {Ising} models and phase
  transitions without local order parameters}.
\newblock \emph{\bibinfo{journal}{Journal of Mathematical Physics}}
  \textbf{\bibinfo{volume}{12}}, \bibinfo{pages}{2259--2272}
  (\bibinfo{year}{1971}).

\bibitem{KogutReview}
\bibinfo{author}{Kogut, J.~B.}
\newblock \bibinfo{title}{An introduction to lattice gauge theory and spin
  systems}.
\newblock \emph{\bibinfo{journal}{Rev. Mod. Phys.}}
  \textbf{\bibinfo{volume}{51}}, \bibinfo{pages}{659--713}
  (\bibinfo{year}{1979}).

\bibitem{SSHModel}
\bibinfo{author}{Su, W.~P.}, \bibinfo{author}{Schrieffer, J.~R.} \&
  \bibinfo{author}{Heeger, A.~J.}
\newblock \bibinfo{title}{Solitons in polyacetylene}.
\newblock \emph{\bibinfo{journal}{Phys. Rev. Lett.}}
  \textbf{\bibinfo{volume}{42}}, \bibinfo{pages}{1698--1701}
  (\bibinfo{year}{1979}).

\bibitem{KitaevModel}
\bibinfo{author}{Kitaev, A.~Y.}
\newblock \bibinfo{title}{Unpaired {Majorana} fermions in quantum wires}.
\newblock \emph{\bibinfo{journal}{Physics-Uspekhi}}
  \textbf{\bibinfo{volume}{44}}, \bibinfo{pages}{131} (\bibinfo{year}{2001}).

\bibitem{TcBKT}
\bibinfo{author}{Komura, Y.} \& \bibinfo{author}{Okabe, Y.}
\newblock \bibinfo{title}{Large-scale {Monte Carlo} simulation of
  two-dimensional classical xy model using multiple {GPUs}}.
\newblock \emph{\bibinfo{journal}{Journal of the Physical Society of Japan}}
  \textbf{\bibinfo{volume}{81}}, \bibinfo{pages}{113001}
  (\bibinfo{year}{2012}).

\bibitem{kPCA}
\bibinfo{author}{Sch\"olkopf, B.}, \bibinfo{author}{Smola, A.} \&
  \bibinfo{author}{M\"uller, K.-R.}
\newblock \bibinfo{title}{Nonlinear component analysis as a kernel eigenvalue
  problem}.
\newblock \emph{\bibinfo{journal}{Neural Computation}}
  \textbf{\bibinfo{volume}{10}}, \bibinfo{pages}{1299--1319}
  (\bibinfo{year}{1998}).

\end{thebibliography}

\vspace{1em}

\sect{Acknowledgements}We thank Shubhayu Chatterjee, Eugene Demler, Peter P.~Orth, Hannes Pichler, Subir Sachdev, Henry Shackleton, and Yi-Zhuang You for useful discussions. JFRN acknowledges support from AFOSR-MURI: Photonic Quantum Matter (award FA95501610323). MSS acknowledges support from the German National Academy of Sciences Leopoldina through grant LPDS 2016-12 and from the National Science Foundation under Grant No.~DMR-1664842.

\sect{Author contributions}JFRN and MSS contributed equally to the design of the study, performing the simulations, analysis of the results, and writing of the manuscript. 

\sect{Competing financial interests}The authors declare no competing financial interests.


\clearpage

\renewcommand{\thefigure}{S\arabic{figure}}
\renewcommand{\theequation}{S\arabic{equation}}
\renewcommand{\thesection}{S\arabic{section}}
\setcounter{page}{1}
\setcounter{equation}{0}
\setcounter{figure}{0}
\setcounter{section}{0}

\begin{widetext}

\begin{center}
{\large\bf Supplementary Information for \\ ``Identifying topological order through unsupervised machine learning"}

\vspace{4mm}

by Joaquin F.~Rodriguez-Nieva and Mathias S.~Scheurer

{\small\it Department of Physics, Harvard University, Cambridge, MA 02138, USA}

\end{center}


\vspace{6mm}

The outline of the Supplement is as follows. In \SD 1, the fundamental differences between diffusion maps and Principal Component Analysis (PCA), a commonly used unsupervised machine learning algorithm, are discussed. In \SD 2, we apply diffusion maps on a more general data set than the one used in the main text for the winding number and compare its performance with PCA. In \SD 3, we derive analytic expressions for the eigenvalues and eigenvectors of the leading order components of diffusion maps. In \SD 4, diffusion maps is applied to a more general data set for the 2D XY model than that of the main text. In \SD 5, we extend the results of Fig.~3(c) of the main text for more values of the temperature $T$. In \SD 6, we show the output of diffusion maps as a function of the parameter $\epsilon$ and quantify the error bars for the critical temperature $T_{\rm c}$. 

\subsection{Supplementary Discussion 1: \\ Fundamental distinctions between diffusion maps and PCA}

Before comparing the performance of diffusion maps and PCA on samples corresponding to topologically distinct mappings, we first discuss the key fundamental differences between the two methods. PCA, or variations of this method such as Kernel PCA, rely on (statistics of) the Euclidean distance between data points and, therefore, are not suitable for topological classification. 
Two samples that belong to the same topological sector can (and generally will)  
have very little overlap with respect to the Euclidean distance. In contrast, diffusion maps captures  
the connectivity of data points with diffusion playing the role of homotopy between configurations.

To illustrate this point in a minimal example, we consider a two dimensional data set comprised of points randomly distributed on two concentric circles, see Fig.~\ref{fig:pcavsdiff}(a). Two antipodal points on either of the two circles, although being adiabatically connected, have large Euclidean distance, which is why they are mapped to very distant points upon application of PCA [Fig.~\ref{fig:pcavsdiff}(b)]. In contrast, the output of diffusion maps [Fig.~\ref{fig:pcavsdiff}(c)] clusters all samples belonging to the same circle on (approximately) the same point since all these points are adiabatically connected.

\begin{figure}[t]
  \centering\includegraphics[scale=1.15]{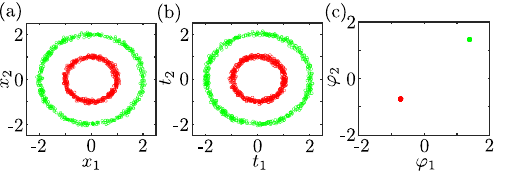}
  \caption{{\bf Minimal example to illustrate difference between diffusion maps and PCA.} (a) Data set comprised of points in a two dimensional plane distributed randomly on two concentric circles. (b) Relying on the Euclidean distance, PCA fails to identify any ordering in the data. (c) Exploiting the connectivity of datapoints using the Euclidean distance, diffusion maps successfully separates the data points by connectivity in each circle. [parameter values: $m=1000$ points, $\epsilon = 0.01$]}
  \label{fig:pcavsdiff}
\end{figure}

\subsection{Supplementary Discussion 2: \\ Winding numbers and diffusion maps}
For pedagogical reasons, we have focused on the simplest scenario of only two winding numbers when discussing the mapping $S_1 \rightarrow S_1$ in the main text. Here, we will extend the data set to show that our scheme based on diffusion maps also works in more general situations and compare it with PCA and kernel PCA.

\subsubsection{Extended data set}
On top of increasing the range of allowed winding numbers $\nu$, we will also extend the form of $\theta_i^{(l)}$ in Eq.~(6) of the main text for the spatial dependence of the angles  of the spins: 
\begin{equation}
    \theta_i^{(l)} = 2\pi \nu^{(l)} i/N + \delta\theta_i^{(l)} +\bar{\theta}^{(l)} + \eta^{(l)} \left[1 -  \cos(2\pi i/N) \right]. \label{ExtendedSetOfAngles}
\end{equation}
As before, $\delta\theta_i^{(l)}$ represents random spin fluctuations which are taken to be Gaussian with standard deviation $\sigma_\theta$, $\bar{\theta}^{(l)}$, which is chosen uniformly between $0$ and $2\pi$, describes global rotations of the spins, and $\nu^{(l)}$ is the integer-valued winding number of sample $l$, which we now draw randomly from the $7$ different values $\{-3,-2,-1,0,1,2,3\}$. The last term in \equref{ExtendedSetOfAngles} has a random prefactor $\eta^{(l)} \in [-\eta_0,\eta_0]$ and further distorts the spin configurations without changing the winding number.

\begin{figure}[t]
  \centering\includegraphics[width=0.95\textwidth]{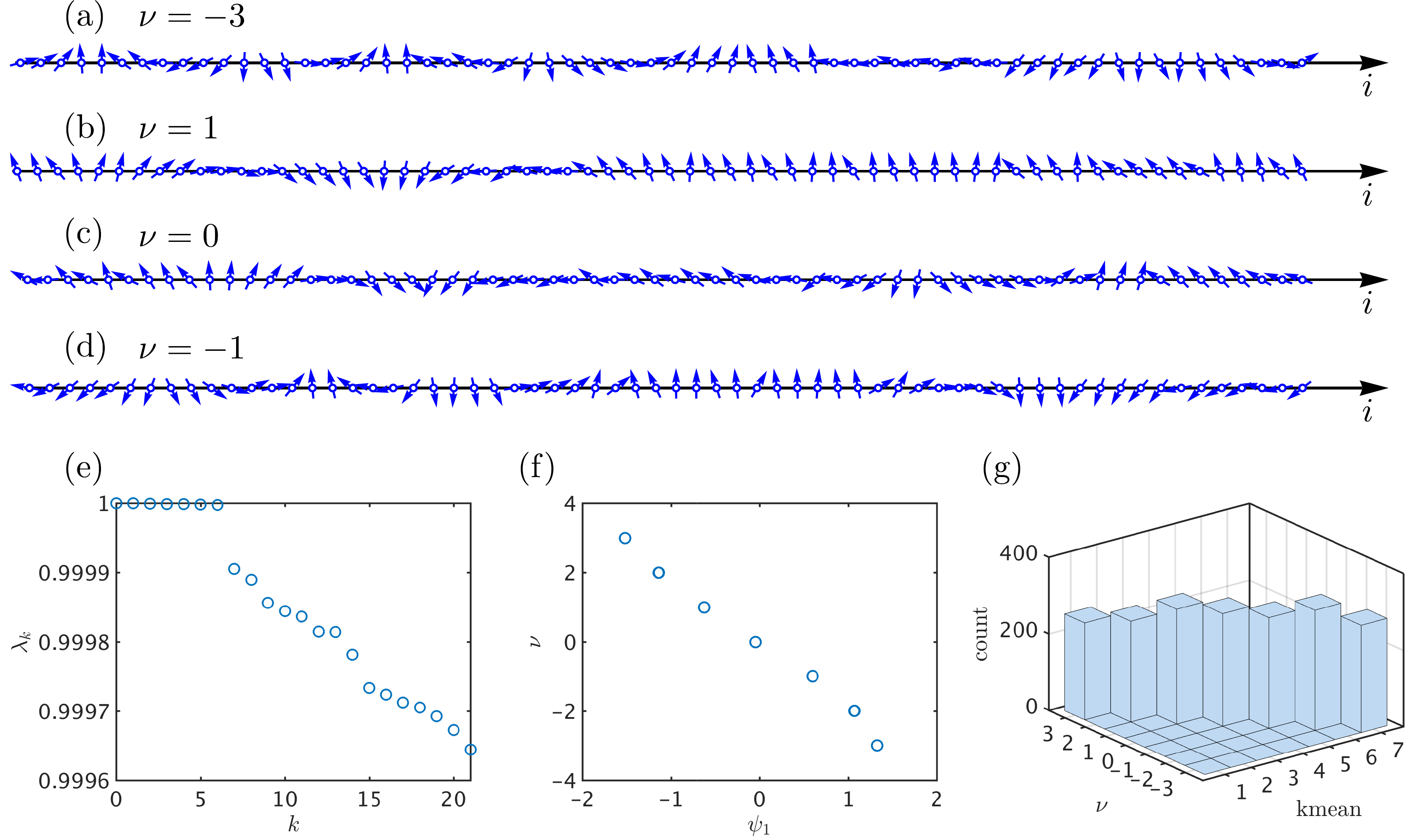}
  \caption{{\bf Performance of diffusion maps on the extended data set.} (a)--(d) show four different typical samples from our training set [see Eq.(\ref{ExtendedSetOfAngles}) for definition]. For clarity, we only indicate the spin of every fourth site. The spectrum of the diffusion operator is shown in (e) revealing the presence of  $7$ sectors in the data set. In (f), we plot the actual winding number $\nu$ as a function of the first component of the diffusion map for all samples (the clustering is so strong that all points in each cluster appear on top of each other in the plot). The method not only clusters samples of the same sector but also discovers the natural ordering of the sectors. The histogram in (g) of the output of $k$-means applied to the reduced ($6$-dimensional) feature space $\vec{\varphi}_l$ reveals that the clustering works with 100\% fidelity. For the diffusion map, we use $\epsilon=0.03$.}
  \label{fig:DMExtendedData}
\end{figure}

We note in passing that these samples can either be viewed as classical spins on a one-dimensional manifold, but alternatively as parametrizing the Hamiltonian $H$ of a two-band model of a 1D topological insulator of symmetry class AIII: the latter can be written as $H(k) = \vec{g}(k) \cdot (\sigma_x,\sigma_y)$, with Pauli matrices $\sigma_{x,y}$. Since, we are interested in gapped phases only, we can focus on $\hat{\vec{g}}(k)=\vec{g}(k)/|\vec{g}(k)|$ which is, again, a mapping from $S_1$ to $S_1$ characterized by a winding number $\nu$; this winding number is the associated topological band index of the insulator. Using samples of $\hat{\vec{g}}(k)$ (discretized in momentum space) as input, $\nu$ has been ``learned'' in \refcite{S1S1MappingSupervised} by a neural network using a supervised approach. Viewing the site index $i$ in \equref{ExtendedSetOfAngles} as discretized momentum, our analysis shows that diffusion maps can learn the topological band index without supervision.

We test the performance of diffusion maps on the extended data set in \equref{ExtendedSetOfAngles} using $\sigma_\theta=0.3$, $\eta_0=4$, $N=256$ sites, and $m=2000$ samples, i.e., about $286$ samples per topological sector. Typical samples of this training set are illustrated in \figref{fig:DMExtendedData}(a)--(d); we notice that, in addition to the high winding numbers [see \figref{fig:DMExtendedData}(a)], the large value of $\eta_0$ leads to strong deformations: e.g., a non-trivial winding can be concentrated in a small part of the chain while the orientation of the spins in the remainder of the system is constant [\figref{fig:DMExtendedData}(b)]. Furthermore, there can be locally non-trivial windings that either completely [\figref{fig:DMExtendedData}(c)] or partially [\figref{fig:DMExtendedData}(d)] ``unwinded'' in other parts of the system.

All of these additional features make the identification of the topological invariant from bare spin configurations a highly non-trivial task. Nonetheless, our approach based on diffusion maps still works: as can be seen in \figref{fig:DMExtendedData}(e), the largest eigenvalue is $7$-fold degenerate, in agreement with the $7$ different sectors. In fact, already the first component, $\psi_1$, of the diffusion map separates all samples according to the winding number $\nu$ [\figref{fig:DMExtendedData}(f)]. On top of just clustering sectors, the approach also learns that there is a natural order of the different sectors as there is a monotonic relation between $\nu$ and $\psi_1$. This results from the fact that the Euclidean distance in the low-dimensional feature space extracted from the diffusion map is an approximation to the diffusion distance and, hence, samples which differ by a small number of windings should be closer than samples that differ by many windings. For completeness, we have also applied $k$-means to the reduced feature space $\vec{\varphi}_l=[(\psi_1)_l,...,(\psi_6)_l]$ and find that all sectors are identified with 100\% fidelity [\figref{fig:DMExtendedData}(g)].

\subsubsection{Comparison to PCA and kernel PCA}

To illustrate that the classification of samples according to winding number is a challenging task for other unsupervised machine-learning algorithms and to clarify the key difference to diffusion maps, we now apply principal component analysis (PCA) and its non-linear extension, kernel PCA \cite{kPCA}.

For simplicity, we will focus on only two different topological sectors, $\nu^{(l)}=\{0,1\}$, take $m=1000$ samples, i.e., about 500 samples per sector, and $\sigma_\theta=0.3$. Let us begin with the simple data set of the main text and set $\eta^{(l)}=0$ in \equref{ExtendedSetOfAngles}. The spectrum of PCA and the projection $t_k^{(l)}$ of the data on the three leading principal components, $k=1,2,3$, are shown in \figref{fig:PCAsSimpleData}(a) and (b), respectively. 
We notice four dominant eigenvalues corresponding to the global rotations in the two topological sectors. While the $t_k^{(l)}$ make the connectivity of the data apparent to the human eye, we see clearly that some pairs of samples with different winding number are still closer to one another in the reduced feature space than pairs of samples with the same $\nu$. This is why $k$-means applied to the four-dimensional feature space $(t_1,t_2,t_3,t_4)$ fails [\figref{fig:PCAsSimpleData}(c)]. Instead an algorithm based on connectivity, as provided by diffusion maps, is required to learn the winding number. As we have shown in the main text and above, diffusion maps can be directly applied to the raw data and does not require the pre-processing step of first performing a PCA. 

\begin{figure}[t]
  \centering\includegraphics[width=\textwidth]{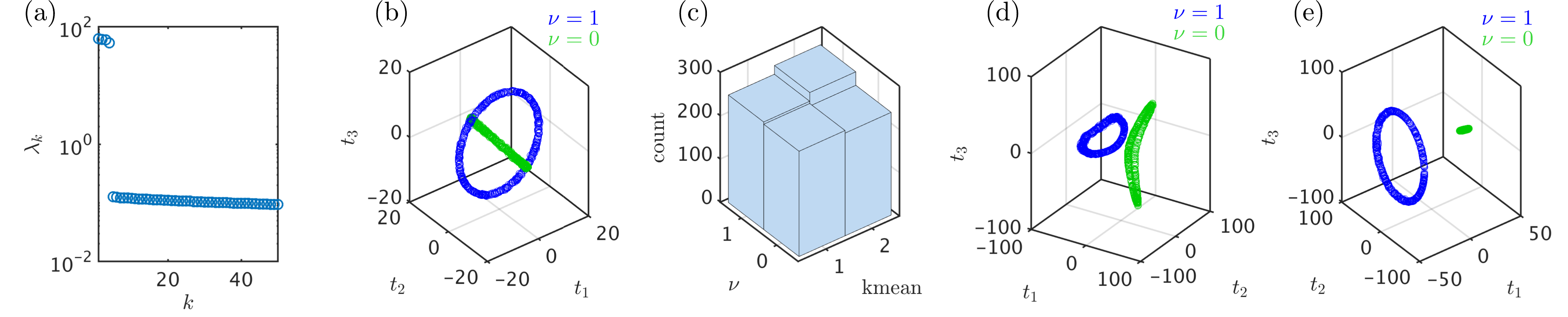}
  \caption{{\bf PCA and kernel PCA on simple data set.} We use a minimally complex data set of the same form as that of the main text, $\eta_0=0$ in \equref{ExtendedSetOfAngles} and with just two distinct winding numbers $\nu=0$, $\nu=1$. We further take $N=256$, $m=1000$, and $\sigma_{\theta}=0.3$. The spectrum, the projections $t_j$ on the three leading principal components, and the statistics of $k$-means applied to the four-dimensional feature space $t_j$, $j=1,2,3,4$, for regular PCA are shown in (a), (b), and (c), respectively. The projection of the data on the low-dimensional feature space of kernel PCA can be found in (d) for a Gaussian kernel $K^g_{35}$ ($\epsilon\approx 35$ seems to work best here) and in (e) for the polynomial kernel $K^p_{2}$.}
  \label{fig:PCAsSimpleData}
\end{figure}

Being linear, it is not surprising that PCA cannot cluster samples by winding number and, hence, we have also applied kernel PCA. \figref{fig:PCAsSimpleData}(d) shows the projection of the data on the three leading components using a Gaussian kernel, $\mathcal{K}^{g}_\epsilon(x,y) = \exp(-||x-y ||^2/(2\epsilon))$, and we find that the same problem persists. 

Due to the simple structure of the data we consider, one might expect that there is a kernel that can separate the data more strongly than the Gaussian kernel. Indeed, the polynomial kernel $\mathcal{K}^p_d(x,y)=(1+x\cdot y)^{d}$, $d\in\mathbb{N}^+$, manages to separate the data much better as can be seen \figref{fig:PCAsSimpleData}(e) (note that $k$-means only works here if we focus on $t_1$). However, kernel PCA has two serious drawbacks: first, its success, even for this simple data set, depends on the correct choice of kernel appropriate to the data set under consideration. Second, and far more importantly, it fails completely for more general data sets as we show next.

Let us now take $\eta_0=4$ exactly as used for the diffusion map above but still focus on $\nu^{(l)}=\{0,1\}$ for simplicity. As can be seen in \figref{fig:PCAsExtendedData}, the information about distinct topological sectors is lost in the low-dimensional feature space of PCA and kernel PCA for both Gaussian and polynomial kernels.

\begin{figure}[t]
  \centering\includegraphics[width=\textwidth]{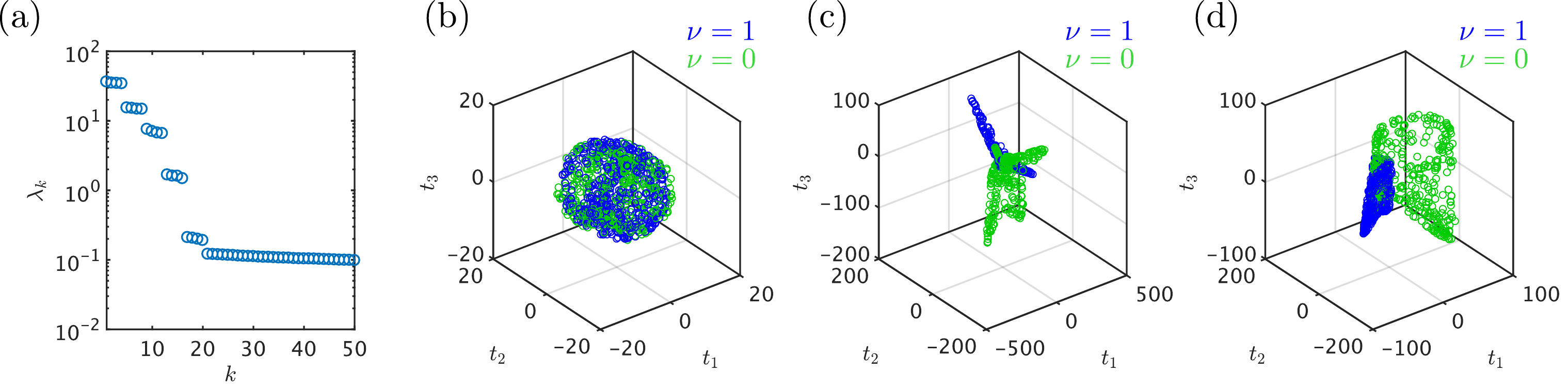}
  \caption{{\bf PCA and kernel PCA for the generalized data set.} Here we use the extended data set with non-zero $\eta_0=4$. The only difference to the data set of \figref{fig:DMExtendedData} is that we restrict the data to two topological sectors and take $m=1000$, i.e., about $500$ samples per sector. The spectrum of PCA and the projected data are shown in (a) and (b). We have also checked that kernel PCA with Gaussian or polynomial kernel cannot separate topological sectors. The associated projection of the data is shown in (c) and (d), obtained for $\epsilon=25$ and $d=4$, respectively.}
  \label{fig:PCAsExtendedData}
\end{figure}

\newpage
\subsection{Supplementary Discussion 3: \\ Splitting between two topological sectors} 

In this section, we derive an expression for the splitting and wavefunctions of the leading (quasi)degenerate modes of $P_{l,l'}$ with $\lambda_k \approx 1$ (i.e., of the zero modes in the language of the quantum well problem). We discuss under which assumptions the splitting is exponentially small and show that the algorithm is robust against an imbalance in the number of samples in different topological sectors. 

For simplicity, let us focus on two topologically distinct sectors, denoted by $A$ and $B$, respectively. This assumes low $T$ such that the topological invariant is well-defined. 
As illustrated in Fig.~2(c) of the main text, the kernel $K_\epsilon(x_l,x_{l'})$ has matrix elements of ${\cal O}(1)$ 
in the blocks with $l,l'\in A$ and  $l,l'\in B$, but is of order $e^{-c/\epsilon}$ if $l \in A$, $l' \in B$ (or vice versa). The value of the order-$1$ constant $c$ in the exponent depends on how strongly the two different sectors are separated with respect to the Euclidean norm in the definition of the kernel in Eq.~(1). 

As mentioned in the main text, we obtain the spectrum of $P_{l,l'}$ by first diagonalizing the matrix 
\begin{equation}
A_{l,l'} = P_{l,l'}\sqrt{\frac{z_l}{z_{l'}}} = \frac{K_{\epsilon}(x_l,x_{l'})}{\sqrt{z_l}\sqrt{z_{l'}}}, \label{DefinitionOfA}
\end{equation}
which has the advantage of being symmetric. Being similar to $P$, it has the same spectrum as $P$ and its eigenfunctions, denoted by $\bar{\psi}_k$ in the following, are related to $\psi_k$ according to $(\psi_k)_l =  (\bar{\psi}_k)_l/\sqrt{z_{l}}$.
We now split the matrix $A$ in \equref{DefinitionOfA} into two parts, $A_{l,l'} = A^{(0)}_{l,l'} + A^{(1)}_{l,l'}$, where $A^{(0)}_{l,l'}$ is nonzero only if $l$ and $l'$ belong to the same topological sector. The matrix $A^{(0)}_{l,l'}$ has an eigenstate with eigenvalue of exactly one in each topological sector. The second part, and $A^{(1)}_{l,l'} = \mathcal{O}(e^{-c/\epsilon})$, will be treated as a perturbation which
will lift the exact degeneracy.

To this end, we use $s\in \{A,B\}$ to label the two sectors and introduce the $\bar{s}$ notation to indicate $\bar{s} = B$ for $s=A$ (and vice versa). For $l\in s$, we write
\begin{equation}
z_l = z_l^s + \Delta z_l^s, 
\end{equation}
where
\begin{subequations}\begin{align}
z_l^s &= \sum_{l'\in s} K_\epsilon(x_l,x_{l'}) = \mathcal{O}(1), \\ 
\Delta z_l^s &= \sum_{l'\in \bar{s}} K_\epsilon(x_l,x_{l'}) = \mathcal{O}(e^{-c/\epsilon}).
\end{align}\end{subequations}
Within this notation, we have
\begin{equation}
A^{(0)}_{l,l'} = \frac{K_{\epsilon}(x_l,x_{l'})}{\sqrt{z^s_l}\sqrt{z^s_{l'}}}, \quad \text{ for } l,l' \in s,
\end{equation}
and $A^{(0)}_{l,l'} = 0$ if $l,l'$ belong to different sectors. This matrix has an eigenvector with eigenvalues $1$ for each sector $s$ which is given by
\begin{align}
(\bar{\psi^s})_l = \begin{cases} \frac{\sqrt{z^s_l}}{\sqrt{\sum_{l'\in s} z^s_{l'}}}, \quad &l \in s, \\ 0, \quad &l \notin s. \end{cases} \label{UnperturbedStates}
\end{align}
The perturbation, $A^{(1)}$, consists of two parts --- one which acts between different sectors, and a contribution resulting from the correction $\Delta z_l^s$ to $z^s_l$. To leading order in $e^{-c/\epsilon}$, we find
\begin{align}
A^{(1)}_{l,l'} \sim \begin{cases} -\frac{K_\epsilon(x_l,x_{l'})}{2\sqrt{z^s_l}\sqrt{z^s_{l'}}}\left( \frac{\Delta z_l^s}{z^s_{l}} + \frac{\Delta z_{l'}^s}{z^s_{l'}} \right), \,\, & l,l' \in s,   \\ \frac{K_\epsilon(x_l,x_{l'})}{\sqrt{z^s_l}\sqrt{z^{\bar{s}}_{l'}}}, \,\, & l \in s, l'\in \bar{s}. \end{cases} \label{PerturbationMatrix}
\end{align}
Given the explicit form of the unperturbed states in \equref{UnperturbedStates} and of the perturbation (\ref{PerturbationMatrix}), it is straightforward to perform degenerate perturbation theory. One finds
\begin{subequations}\begin{align}
\lambda_0 &= 1, \\ 
\begin{split}\lambda_1 & = 1 - M_{AB} \left( \frac{1}{M_{AA}} + \frac{1}{M_{BB}}\right ) + \mathcal{O}(e^{-2c/\epsilon}), \label{ReductionFrom1}\end{split}
\end{align}\end{subequations}
where we have defined
\begin{equation}
M_{ss'} = \sum_{l\in s}\sum_{l' \in s'} K_{\epsilon}(x_l,x_{l'}).
\end{equation}
The associated wavefunctions read as
\begin{subequations}\begin{align}
\bar{\psi}_0 &= \mathcal{N} \left( \eta \, \bar{\psi}^A + \bar{\psi}^B \right), \quad \eta = \sqrt{M_{AA}/M_{BB}}, \\
\bar{\psi}_1 &= \mathcal{N} \left( \bar{\psi}^A - \eta \, \bar{\psi}^B \right) 
\end{align}\end{subequations}
with normalization factor $\mathcal{N}$.

First, note that the larger eigenvalue, $\lambda_0$, is still $1$ which, in fact, must hold to all orders in perturbation theory as a result of global probability conservation, $\sum_{l'} P_{l,l'}=1$. As expected, the associated eigenstate, $\bar{\psi}_0$, is the symmetric superposition of the two eigenstates, $\bar{\psi}^A$ and $\bar{\psi}^B$, found by first considering the topologically distinct sectors separately. 
Because $M_{AB} = \mathcal{O}(e^{-c/\epsilon})$, the eigenvalue of the antisymmetric combination $\bar{\psi}_1$ is reduced from one by an exponentially small amount, see \equref{ReductionFrom1}. More precisely, this splitting scales as
\begin{equation}
1- \lambda_1 \propto e^{-c/\epsilon} \left( \frac{m_A}{m_B} + \frac{m_B}{m_A}\right) \label{ScalingOfEigenvalue}
\end{equation}
with $\epsilon$ and the number of samples $m_s$ in sector $s$.  Consequently, as long as the asymmetry ratio $R=\max(\frac{m_A}{m_B},\frac{m_B}{m_A})$ is much smaller than the exponential scale $e^{c/\epsilon}$, the perturbative approach presented here is valid. In this case, the quasi-degenerate eigenstates should be clearly visible in the spectrum of the diffusion map, and deviations from degeneracy scale exponentially with $\epsilon$.

We finally mention that the weak requirement $R \ll e^{c/\epsilon}$ can be further relaxed by proper rescaling of the kernel. Using (this corresponds to $\alpha=1$ in the notation of \refcite{COIFMAN20065}) the kernel 
\begin{subequations}\begin{align}
K_{\epsilon}(x_l,x_{l'}) &= \frac{\exp\left(-\frac{||x_l-x_{l'}||^2}{2 N \epsilon}\right)}{n_\epsilon(x_l)n_\epsilon(x_{l'})}, \\
n_\epsilon(x_l) &= \sum_{l'} \exp\left(-\frac{||x_l-x_{l'}||^2}{2 N \epsilon}\right), 
\end{align}\label{NewKernel}\end{subequations}
instead of the kernel in Eq.~(1), the factors $m_A/m_B$ and $m_B/m_A$ in the scaling (\ref{ScalingOfEigenvalue}) of the eigenvalue will be eliminated at order $e^{-c/\epsilon}$. 
We have checked that the BKT transition can be equally well detected (yielding the same $T_c$ within the accuracy of our simulations) with the kernel in \equref{NewKernel}.

\subsection{Supplementary Discussion 4: \\ Generalized data set for the 2D XY model}

In the main text, we used diffusion maps for the five most energetically favorable topological sectors and sampled each sector equally. The linear length of the samples was $L = 32$. Here we extend our results to include more general data sets and demonstrate that diffusion maps successfully classify samples in each of these cases. 

\subsubsection{Clustering for random winding numbers and uneven sampling}

We first demonstrate that diffusion maps can correctly classify topological sectors for arbitrary winding numbers and uneven sampling in each topological sector. Because the system is finite, we restrict the possible $\nu_{x,y}$ to a finite range $|\nu_{x,y}|\le 3$ for a system of linear length $L=32$ (states with $\nu_{x,y}\sim L$ are energetically unstable). Further, we create 500 samples for the first topological sector, 1000 samples for the second topological sector, and 2000 samples for the third topological sector. We use a moderately large temperature $T/J = 0.45 \sim T_{\rm c}/2$ to create our samples. Figure~\ref{fig:randomwinding}(a) shows $\lambda_{k<10}$ corresponding to the output of diffusion maps. The presence of a three fold degeneracy for $\lambda_k =1$ signals the presence of three topological sectors in our samples. By plotting the first two nontrivial eigenvectors $\psi_{1,2}$, we observe clustering of the samples into three topological sectors with the correct labels assigned to each of them. 
\begin{figure}
  \centering\includegraphics[scale=1.15]{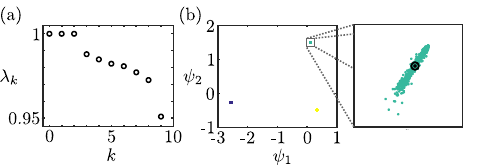}
  \caption{{\bf Diffusion maps for random winding numbers and uneven sampling in each sector}. (a) First ten eigenvalues of the diffusion map exhibiting a three fold degeneracy of $\lambda_k =1$. (b) Two dimensional feature space $\psi_1$ and $\psi_2$ showing three topological sectors (axes were normalized with intercluster distance $\bar D$). Colorcode indicates the winding number of the samples. The right panel shows a zoom of one of the topological sectors.}
  \label{fig:randomwinding}
\end{figure}

\subsubsection{Diffusion maps as a function of system size}

Here we check the success of diffusion maps for different system sizes. Figure~\ref{fig:systemsize} shows $\bar\sigma$ and $\bar{D}$ for the 2D XY model of system size $L = 48$, and $L = 64$ using the same winding numbers as in the main text. We find qualitatively the same results as in the $L = 32$ case used in the main text. In particular, for $T/J < 0.8$, the algorithm can correctly classify samples by topological sector, i.e. $\bar\sigma \ll \bar{D}$, independently of system size. For $T/J > 1$, the algorithm only finds one cluster. We define the phase transition point as the point in which $\frac{2\bar\sigma}{\bar D} = \frac{1}{n}$. By checking the sensitivity of the results as a function of $\epsilon$, see below, we find that the phase transition occurs in the range $T_{\rm c}/J = 0.9\pm 0.1$. 

\begin{figure}[t]
  \centering\includegraphics[scale=1.0]{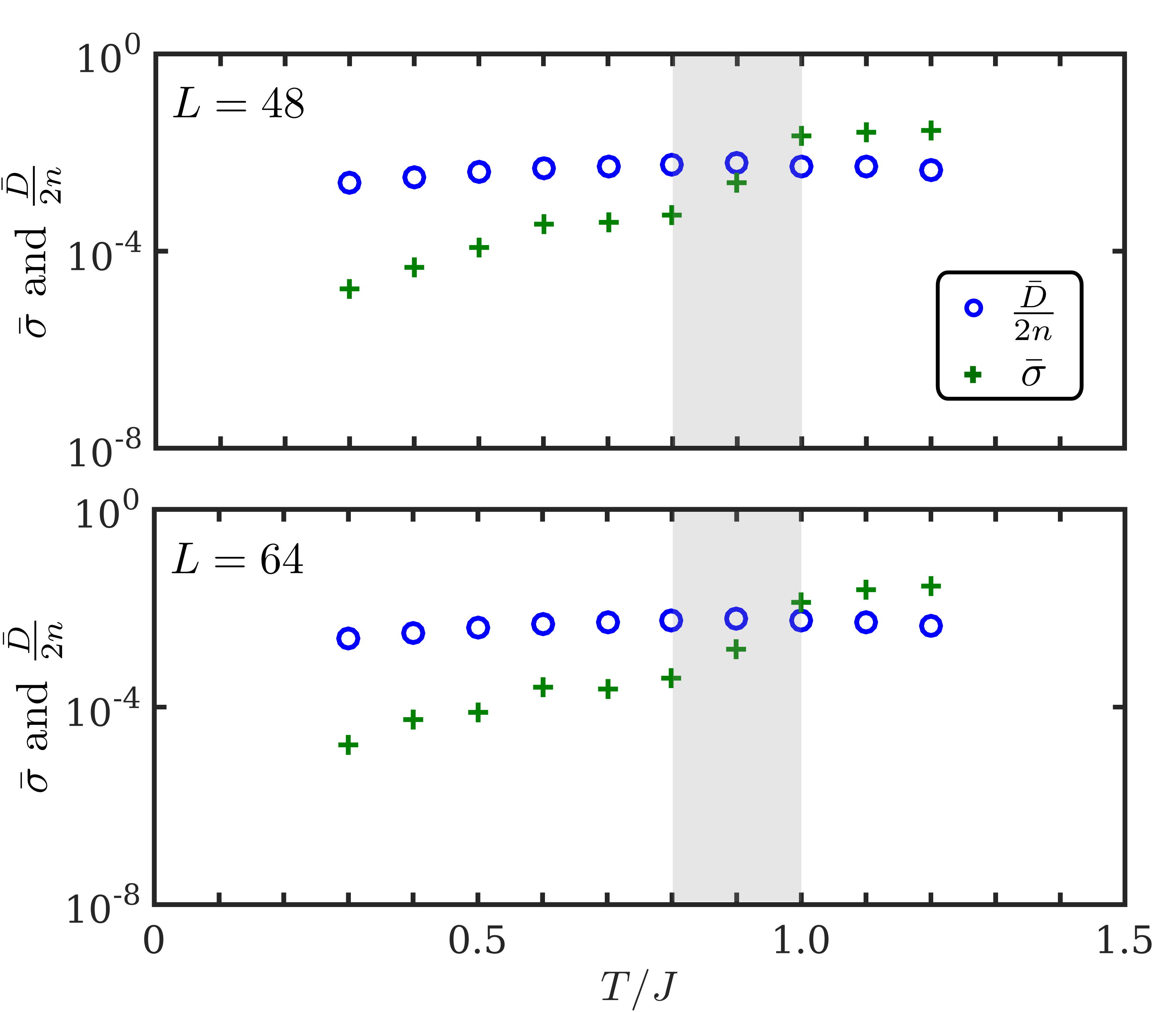}
  \caption{{\bf Success of the clustering algorithm as a function of system size}. Here we extend  the results of Fig.~3(d) in the main text for the   2D XY model with system length (a) $L=48$ and (b) $L=64$, $\epsilon/\epsilon_* = 5$. }
  \label{fig:systemsize}
\end{figure}

\begin{figure}[t]
  \centering\includegraphics[scale=1.0]{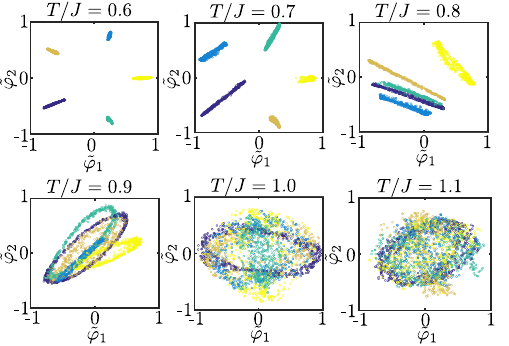}
  \caption{{\bf Clustering of the 2D XY samples close to the topological phase transition.} Shown is the projection of the 4-dimensional reduced feature space, Eq.~(7) of the main text, on a two-dimensional plane for increasing values of $T/J$. The color code refers to the winding number of each sample.}
  \label{fig:clusteringfull}
\end{figure}

\newpage

\begin{figure}
  \centering\includegraphics[scale=1.0]{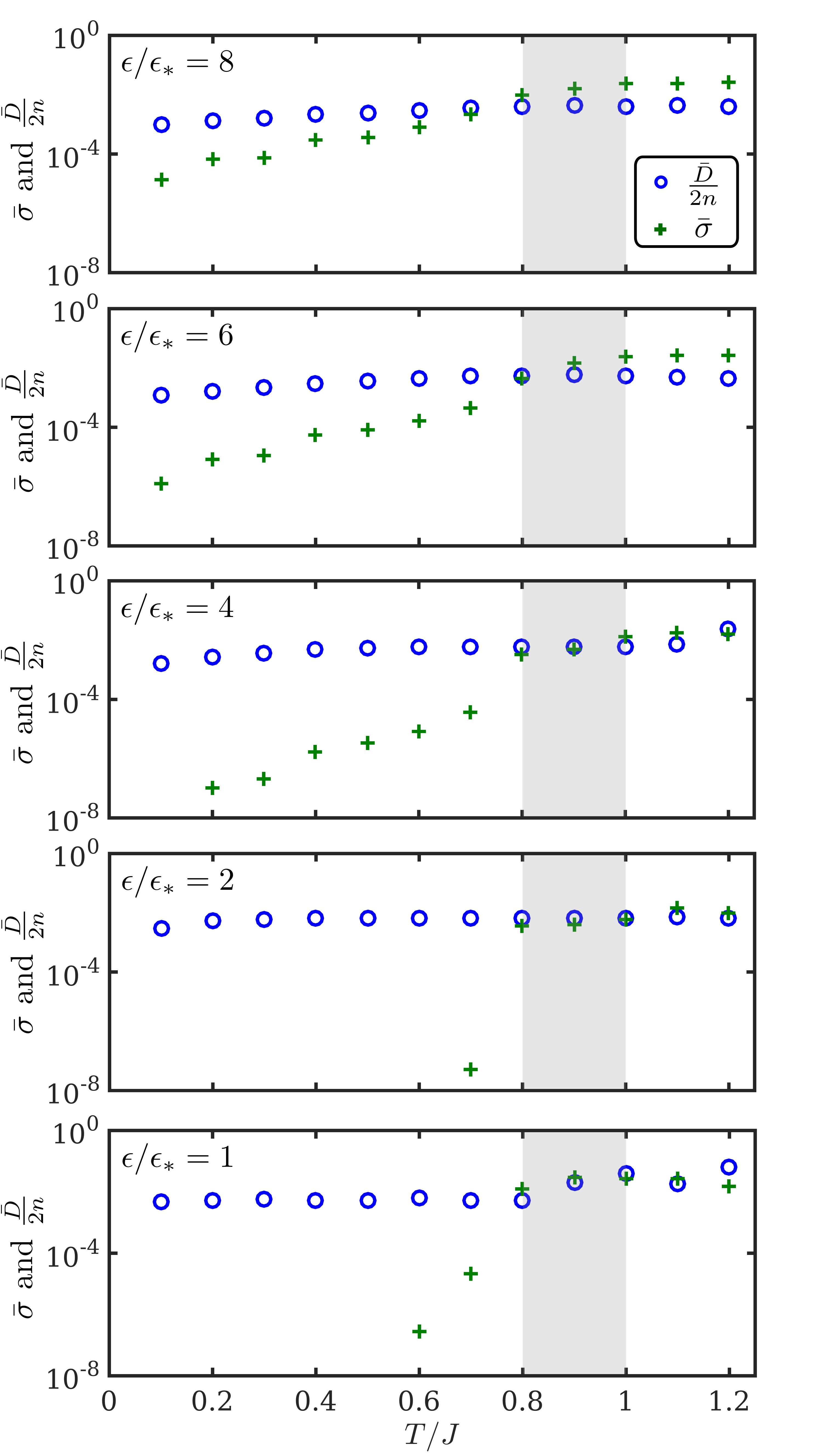}
  \caption{{\bf Success of diffusion maps for varying $\epsilon$}. For $T/J\lesssim 0.7$, diffusion maps can clearly diagnose topological sectors given that   $\bar{\sigma}\ll \bar{D}$ regardless of $\epsilon$. We defined the transition temperature as the temperature in which $\frac{2\bar{\sigma}}{\bar D}= \frac{1}{n}$. The transition temperature varies with the choice of $\epsilon$, and on average is given by   $T/J = 0.9 \pm 0.1$, see shaded area (datapoints $\bar\sigma < 10^{-8}$ are not shown). [$m=2500$ evenly distributed in five topological sectors]}
  \label{fig:epsilondependence}
\end{figure}

\subsection{Supplementary Discussion 5: \\ Clustering across the BKT transition}

In this section, we extend the results of Fig.~3(d) of the main text to illustrate in more detail how the clustering of the samples fails as the phase transition is approached. Figure~\ref{fig:clusteringfull} shows the projection of the low-dimensional feature space, Eq.~(7) in the main text, into a two-dimensional plane that has been optimized to make the clustering clearly visible. For $T/J\le 0.6$, samples are tightly clustered by topological sector. In the range, $T/J = 0.7{\rm-}0.8$, clusters are still well defined, but they start to merge. For $T/J\ge 0.9$, the clusters merge into a single cluster; topological sectors are no longer well defined, which is the defining feature of the topological phase transition. 

\subsection{Supplementary Discussion 6: \\ Sensitivity of diffusion maps to the kernel $K_\epsilon$}

Here we quantify the error bars associated with $T_{\rm c}$ by checking the output of diffusion maps as a function of $\epsilon$ in Eq.~(1) of the main text. Figure~\ref{fig:epsilondependence} shows the values of $\bar{D}$ and $\bar{\sigma}$ for the same datapoints used in Fig.~3(d) of the main text and $\epsilon$ varying in the range $1<\epsilon/\epsilon_* <8 $ ($\epsilon_* = \frac{2\pi}{m_\nu}$, $m_{\nu}$: number of samples per topological sector). For $T/J\lesssim 0.8$, we find that datapoints are correctly clustered by topological sectors, i.e.~$\bar{\sigma}\ll \bar{D}$, regardless of the value of $\epsilon$. For $T/J \gtrsim 1$, the algorithm only finds a single cluster of samples regardless of $\epsilon$. The exact value in which $\frac{2\bar\sigma}{\bar D} = \frac{1}{n}$ depends weakly on $\epsilon$. In particular, we note that $\epsilon$ varies one order of magnitude in Fig.~\ref{fig:epsilondependence}, but $T_{\rm c}$ varies only within $\pm 15\,\%$. We also note that the results do not vary sensitively with the criterion $\bar\sigma/2\bar{D} = 1/n$ (note that the $y$-axis in Fig.~\ref{fig:epsilondependence} is in logarithmic scale). As a result, we estimate the critical temperature to be in the range $T_{\rm c}/J = 0.9 \pm 0.1$ for a wide range of $\epsilon$ values. 

\end{widetext}

\end{document}